\documentclass[prd,nofootinbib,floats,twocolumn,superscriptaddress]{revtex4-1}
\usepackage{slashed,amsmath,amsfonts,amssymb,hyperref,graphicx,bm}
\usepackage[utf8]{inputenc}
\graphicspath{{fig/}}

\usepackage{ulem}
\usepackage{color}
\usepackage{xcolor}
\definecolor{purple}{rgb}{0.8,0,0.6}
\definecolor{orange}{rgb}{1,0.55,0}

\newcommand{\add}[1]{{\color[rgb]{0,0,0}{#1}}}

\newcommand{\eq}[1]{(\ref{#1})}
\newcommand{\beq}{\begin{equation}}
\newcommand{\eeq}{\end{equation}}
\newcommand{\beqn}{\begin{eqnarray}}
\newcommand{\eeqn}{\end{eqnarray}}
\newcommand{\beqs}{\begin{subequations}}
\newcommand{\eeqs}{\end{subequations}\\[-2mm]\noindent}

\newcommand{\cL}{{\mathcal{L}}}

\renewcommand{\P}{{\cal{P}}}
\newcommand{\T}{{\cal{T}}}
\newcommand{\PT}{{\mathcal{PT}}}

\newcommand{\NH}{{\mbox{\tiny{NH}}}}
\newcommand{\HH}{{\mbox{\tiny{H}}}}

\newcommand{\bs}{\boldsymbol}
\newcommand{\x}{{\bs x}}
\newcommand{\Z}{{\mathbb Z}}
\newcommand{\R}{{\mathbb R}}
\def\bbbone{{\mathchoice {\rm 1\mskip-4mu l} {\rm 1\mskip-4mu l} {\rm 1\mskip-4.5mu l} {\rm 1\mskip-5mu l}}}

\begin{document}
\title{\add{The phase diagram and vortex properties} of $\PT$-symmetric non-Hermitian two-component superfluid}

\author{A. M. Begun}
\affiliation{Pacific Quantum Center, Far Eastern Federal University, Sukhanova 8, Vladivostok 690950, Russia}
\author{M. N. Chernodub}
\affiliation{Institut Denis Poisson UMR 7013, Universit\'e de Tours, 37200 France}
\affiliation{Pacific Quantum Center, Far Eastern Federal University, Sukhanova 8, Vladivostok 690950, Russia}
\author{A. V. Molochkov}
\affiliation{Pacific Quantum Center, Far Eastern Federal University, Sukhanova 8, Vladivostok 690950, Russia}

\date{\today}

\begin{abstract}
\add{
We discuss the phase diagram and properties of global vortices in the non-Hermitian parity-time-symmetric relativistic model possessing two interacting scalar complex fields. The phase diagram contains stable $\PT$-symmetric regions and unstable $\PT$-broken regions, which intertwine nontrivially with the U(1)-symmetric and U(1)-broken phases, thus forming rich patterns in the space of parameters of the model. The notion of the $\PT$ symmetry breaking is generalized to the interacting theory. At finite quartic couplings, the non-Hermitian model possesses classical vortex solutions in the $\PT$-symmetric regions characterized by broken U(1) symmetry. In the long-range limit of two-component Bose-Einstein condensates, the vortices from different condensates experience mutual dissipative dynamics unless their cores overlap precisely. For comparison, we also consider a close Hermitian analog of the system and demonstrate that the non-Hermitian two-component model possesses much richer dynamics than its Hermitian counterpart.
}
\end{abstract}

\maketitle

\section{Introduction}

Quantum-mechanical systems are traditionally described by Hermitian Hamiltonians which ensure the real-valuedness of the full energy spectrum and, therefore, the unitary evolution of the system as a whole. It turns out, however, that the Hermitian description can be extended with a large class of non-Hermitian terms which are invariant under combined parity-time ($\PT$) transformations. The $\PT$-symmetric non-Hermitian systems are as meaningful as the conventional Hermitian quantum mechanics in the regions where the $\PT$-symmetry is not broken spontaneously~\cite{Bender:1998ke,Bender:2005tb}.

Mathematically, in the $\PT$-symmetric systems the familiar Hermiticity condition $H^\dagger = H$ is replaced by the requirement of the $\PT$-symmetricity $(\PT) H (\PT) = H$, which is equivalent to the commutation of the Hamiltonian with the combined parity $\P$ and time-reversal inversion $\T$ operation~\cite{Bender:2005tb}, $[\PT,H] = 0$. This combined symmetry leads to real-valued energy spectrum that ensures the stability of the system. One can show that all $\PT$-symmetric non-Hermitian Hamiltonians belong to the class of the so-called pseudo-Hermitian Hamiltonians $\eta H \eta^{-1} = H^\dagger$ where $\eta$ is a Hermitian linear automorphism~\cite{Mostafazadeh:2001jk}. The pseudo-Hermiticity is a generalization of the $\PT$-symmetry which, in turn, depends crucially on the fact that the time-reversal transformation $\T$ is an anti-linear operation. 

As it was shown in Ref.~\cite{ref:Wigner:1960} and most recently emphasized in Ref.~\cite{Mannheim:2015hto}, it is the anti-linearity property rather than Hermiticity which is important for the self-consistent description of stable quantum-mechanical systems. The non-Hermitian $\PT$-invariant quantum systems can be mapped to their Hermitian counterparts via a non-Unitary transformation~\cite{Bender:2005hf,Alexandre:2017foi} (the existence of the map is not guaranteed as there are known exceptions in quantum mechanics~\cite{Bender:2008gh}). These extensions broaden the class of stable physical systems beyond the tight Hermiticity constraints and open new horizons for the research. 

The non-Hermitian description has been extended to interacting relativistic field theories, including the systems of fundamental particle interactions. The $\PT$-symmetric interactions which explicitly break the non-Hermiticity of the system can arise in fermionic theories~\cite{Beygi:2019qab}, contribute to the mass gap generation in the NJL model and affect the phase structure of the model~\cite{Felski:2020vrm}. The non-Hermitian Dirac fermions allow for the realization of an anomalous equilibrium transport~\cite{Chernodub:2019ggz}. The ordinary Hermitian models can generate a new, non-Hermitian ground state which could potentially be formed, for example, in fireballs of quark-gluon plasma created after heavy-ion collisions~\cite{Chernodub:2020cew}. In the context of the extensions of the Standard Model of particle interactions, anti-Hermitian Yukawa interactions may lead to an anomalous radiative mass-gap generation in a model of the right-handed sterile neutrinos~\cite{Alexandre:2020bet,Mavromatos:2020hfy}. The concept of non-Hermitian quantum theory allows an extension via the gauge-gravity duality well beyond the scope of the field-theoretical models~\cite{Arean:2019pom}.

The $\PT$-symmetric non-Hermitian Hamiltonians arise in the description of various open quantum systems in optics and solid state physics where this symmetry can be interpreted as a result of a perfect balance between the gains and losses as the system interacts with the external environment~\cite{ref:Rotter:2009,ref:optics:2010}. The recent works also include the studies of the effect of non-Hermitian terms in topological superconductivity which leads to nonlocal anomalous transport effects~\cite{Kawabata:2018} as well as in the conventional superconductivity which gives rise to the unusual first-order phase transition between the phases~\cite{Ghatak:2018}. The possibility of non-Hermitian superfluidity with a complex-valued, non-Hermitian interaction constant naturally arises from inelastic scattering between fermions~\cite{Yamamoto:2019}. The associated non-Hermitian BCS-BEC crossover of Dirac fermions in field-theoretical models of many-body systems reveals a nontrivial phase diagram as a function of the complex coupling~\cite{Kanazawa:2020amq}. 

In our paper, we work with vortex topological defects in a bosonic non-Hermitian model which involves a pair of scalar fields associated with interacting condensates. The topological solutions in the multicomponent scalar models are interesting because they appear in the models which have applications from condensed matter to high energy physics. Some of these models can serve as viable extensions of the Standard model of fundamental particle physics~\cite{Alexandre:2018xyy,Alexandre:2018uol,Fring:2019hue,Alexandre:2020gah}. Similarly to the Grand Unification particle models and their close counterparts, they host 't~Hooft--Polyakov monopole configurations~\cite{Fring:2020xpi} along with complex skyrmions~\cite{Correa:2021pwi} and kink/anti-kink solutions~\cite{Fring:2021zci} with real-valued energies. As in the Hermitian models, these classical solutions are associated with the saddle points of the corresponding partition functions. 

At the condensed matter side, the many-condensate systems possess richer dynamics than their one-condensate counterparts. 
For example, the standard classification of superconductivity into types I and II fails to describe the phases of multiband condensates so that a proposal to adopt a new terminology, a type-1.5 superconductivity, appeared in the theoretical community~\cite{ref:type1.5:1}. Experimentally, the existence of the type-1.5 superconductivity has been demonstrated shortly afterwards~\cite{ref:type1.5:2}. The semi-Meissner state of a type-1.5 superconductor demonstrates non-pairwise interaction between the vortices which leads to formation of a multitude of complicated vortex states~\cite{ref:type1.5:3}. We discuss the non-Hermitian extension of the two-component model possessing a global, rather than local, continuous symmetry, appropriate for the two-component superfluidity. We concentrate on stability of the ground state, fate of the $\PT$ symmetry in the interacting model, and the properties of the vortex configurations. In a different context, the vortices in the weakly interacting superfluid Bose-Einstein condensates with complex-valued $\PT$-symmetric potentials have been investigated in Ref.~\cite{ref:vortices:1}. \add{As for the realistic model, one could expect the model can apply to the non-Hermitian version of two-component Bose-Einstein condensates in different hyperfine spin states similarly to the Hermitian analogues~\cite{ref:vortices:review}.}

\add{Non-Hermitian two-field models appear, for example, in the description of two-component out-of-equilibrium condensates in a non-Hermitian system of electron-hole pairs and photons in a semiconductor microcavity system. This open quantum many-body system resides in steady-state regimes characterized by a nontrivial phase diagram which contains an exceptional point that marks an endpoint of the first-order phase boundary~\cite{ref:add:one} and exhibits anomalous critical phenomena~\cite{ref:add:two}.}

The plan of our paper is as follows. In Section~\ref{sec:Lagrangians} we briefly overview the Lagrangian and its symmetries, and discuss the ground state of the minimal non-Hermitian theory with two scalar fields. The special attention is paid to the extension of the analysis of the $\PT$ symmetries to the case of interacting model. \add{In particular, we discuss the fate of the $\PT$ symmetry breaking in the presence of interactions}. In Section~\ref{sec:vortices:London} we consider \add{the vortex properties for condensates with frozen radial degrees of freedom for which, in the absence of vortices, the condensate density is approximately uniform and the dynamics is encoded in the phase of the field.} In Section~\ref{sec:vortices:finite} we describe the examples of the vortex solutions at finite quartic couplings. The last section is devoted to our conclusions.

\section{(Non-)Hermitian scalar theory}
\label{sec:Lagrangians}

\subsection{Lagrangians}

We consider a simplest example of a scalar non-Hermitian theory which describes a $\PT$-symmetric dynamics of two complex scalar fields $\phi_1$ and $\phi_2$ conveniently grouped into the single doublet field,
\beqn
\Phi = \begin{pmatrix}
\phi_1 \\ \phi_2
\end{pmatrix}\,.
\eeqn
The Lagrangian of the theory~\cite{Alexandre:2017foi}, 
\beqn
{\cal L} & = & \partial_\mu \Phi^\dagger \partial^\mu \Phi - \Phi^\dagger {\hat M}^2 \Phi - V(\Phi),
\label{eq:L}
\eeqn
includes the classical Hermitian self-interaction potential for the scalar fields:
\beqn
V(\Phi) \equiv V(\phi_1,\phi_2) = \lambda_1 |\phi_1|^4 + \lambda_2 |\phi_2|^4\,.
\label{eq:V}
\eeqn

The non-Hermiticity is encoded in the real-valued mass matrix ${\hat M}^2$ of the Lagrangian~\eq{eq:L}:
\beqn
{\hat M}^2 = {\hat M}^2_\NH = \begin{pmatrix}
\phantom{-} m_1^2 & m_5^2 \\[1mm]
- m_5^2 & m_2^2
\end{pmatrix}\,,
\label{eq:mass:NH}
\eeqn
provided the off-diagonal element\footnote{We use the notation $m_5$ which is also appropriate for the non-Hermitian mass of a fermionic model~\cite{Chernodub:2019ggz}.} of this matrix is a nonzero, $m_5^2 \neq 0$. To see how the non-Hermiticity enters the theory, it is instructive to write the Lagrangian in terms of the individual fields $\phi_1$ and $\phi_2$:
\beqn
{\cal L}_\NH & = & \partial_\nu\phi_1^* \partial^\nu\phi_1+\partial_\nu\phi_2^*\partial^\nu\phi_2  - m_1^2|\phi_1|^2-m_2^2|\phi_2|^2 
\nonumber \\
& & - m_5^2(\phi_1^*\phi_2 - \phi_2^*\phi_1) - \lambda_1 |\phi_1|^4 - \lambda_2 |\phi_2|^4.
\label{eq:L:prime}
\eeqn
The first term of the second line in Eq.~\eq{eq:L:prime} takes a purely complex value: $- 2 i m_5^2 \, {\mathrm{Im}}\, (\phi_1^*\phi_2)$ if the off-diagonal component of the mass matrix~\eq{eq:mass:NH} is a real-valued nonzero quantity. The complex valuedness of the Lagrangian~\eq{eq:L:prime} is consistent with the non-Hermiticity of the mass matrix in Eq.~\eq{eq:L}: ${\hat M}^{2,\dagger}_\NH \neq {\hat M}^2_\NH$. The off-diagonal mass with $m_5^2 \neq 0$ sets up the non-Hermitian regime in \eq{eq:mass:NH}, while the point $m_5^2 = 0$ corresponds to a Hermitian theory (with ${\hat M}^{2,\dagger}_\NH = {\hat M}^2_\NH$) which describes two non-interacting scalar fields $\phi_1$ and $\phi_2$. 

The model~\eq{eq:L} describes two relativistic superfluids which interact with each other via the off-diagonal non-Hermitian coupling. 
We consider the potential~\eq{eq:V} in the form which explicitly breaks the $U(2)$ symmetry, $\Phi \to \Omega \Phi$ with the $2 \times 2$ matrix $\Omega \in U(2)$, down to its Cartan $[U(1)]^2$ subgroup since the $U(2)$ group is explicitly broken by the mass matrix~\eq{eq:mass:NH} anyway provided ${\hat M}^2_\NH \not\propto \bbbone$.

In order to highlight the features of non-Hermiticity, we briefly discuss the Hermitian version of the model with the following mass matrix in the Lagrangian~\eq{eq:L}:
\beqn
{\hat M}^2_\HH = \begin{pmatrix}
m_1^2 & m_5^2 \\[1mm]
m_5^2 & m_2^2
\end{pmatrix}
\label{eq:mass:H}
\eeqn
Notice that ${\hat M}^{2,\dagger}_\HH = {\hat M}^2_\HH$ as expected and the Hermitian Lagrangian is a real-valued expression for all values of the parameter $m_5$:
\beqn
{\cal L}_\HH & = & \partial_\nu\phi_1^* \partial^\nu\phi_1 + \partial_\nu\phi_2^*\partial^\nu\phi_2
 - m_1^2|\phi_1|^2 - m_2^2|\phi_2|^2 \nonumber \\
& & - m_5^2(\phi_1^*\phi_2+\phi_2^*\phi_1) - \lambda_1 |\phi_1|^4 - \lambda_2 |\phi_2|^4\,.
\label{eq:L:H}
\eeqn

\add{
As we discuss in Section~\ref{sec:nonHermitian:vortices}, the Hermitian counterpart~\eq{eq:L:H} of the non-Hermitian system~\eq{eq:L:prime} has some parallels with effective models of the QCD strings in high-density quark matter, relevant to the physics of neutron stars, and the global cosmological strings, which reveal itself in the cosmological context~\cite{ref:A,ref:B,ref:C}.
}

Below, we will consider the classical equations of motion concentrating on the non-Hermitian theory described by Lagrangian~\eq{eq:L}. While we work with the classical solutions, we would like to notice that on quantum level, the non-Hermiticity propagates into the loops of perturbation theory. The quantum corrections could, therefore, induce a complex term in the interaction potential~\eq{eq:V} and literally complexify the phase diagram of the theory. Leaving aside the quantum corrections, which were discussed in Ref.~\cite{Alexandre:2018xyy}, in our paper we concentrate on classical properties of the Lagrangian~\eq{eq:L} \add{and compare it with the classical properties of its Hermitian counterpart~\eq{eq:L:H}.}

\subsection{Symmetries}

We consider the non-degenerate model~\eq{eq:L} with a nonzero off-diagonal mass $m_5 \neq 0$ which possesses a couple of continuous and discrete symmetries. Both Hermitian and non-Hermitian versions of the model are invariant under the global U(1) transformation 
\beqn
U(1): \qquad \Phi(t,\x) \to \Phi'(t,\x) = e^{i \omega} \Phi(t,\x),
\label{eq:U1}
\eeqn
in which the single phase factor (with a real-valued parameter $\omega$) is shared by both complex scalar fields $\phi_1$ and $\phi_2$. If $m_5 = 0$, both components of the doublet field~$\Phi$ can be transformed independently so that the symmetry is enlarged to the global [U(1)]${}^2$. We do not consider this trivial case.

The non-Hermitian theory~\eq{eq:mass:NH} is also invariant under a discrete transformation corresponding to the product of the parity inversion $\P$ and the time conjugation $\T$ operators, respectively:
\beqn
\mbox{NH}: \quad \left\{
\begin{array}{rcl}
\P: & & \Phi(t,\x) \to \Phi'(t,-\x) = \sigma_3 \Phi(t,\x), \quad \\[2mm]
\T: & & \Phi(t,\x) \to \Phi'(-t,\x) = \Phi^*(t,\x).
\end{array}
\right.
\eeqn
Due to the presence of the Pauli matrix $\sigma_3$, the upper $\phi_1$ component transforms under the parity inversion $\P$ as a genuine scalar while the lower component $\phi_2$ transforms as a pseudoscalar field. 

The $\P$ and $\T$ operations whose product leaves invariant the Hermitian theory~\eq{eq:mass:H},
\beqn
\mbox{H}: \quad \left\{
\begin{array}{rcl}
\P: & & \Phi(t,\x) \to \Phi'(t,-\x) = \Phi(t,\x), \quad \\[2mm]
\T: & & \Phi(t,\x) \to \Phi'(-t,\x) = \Phi^*(t,\x),
\end{array}
\right.
\eeqn
indicate that both fields $\phi_1$ and $\phi_2$ behave as true scalars under the parity inversion.

The model~\eq{eq:L}, along with its extensions, possesses interesting features of the Goldstone modes associated with the spontaneous breaking of the continuous symmetry~\eq{eq:U1} as well as the unusual properties of the conserved currents~\cite{Alexandre:2017foi,Mannheim:2018dur,Fring:2019hue}. Below we consider the ground state and the vortex solutions of the model.

\subsection{Ground states}

\subsubsection{Non-Hermitian ground state}

The analysis of the ground state of the two-field model~\eq{eq:L} has already been done analytically in Ref.~\cite{Mannheim:2018dur} for the special case of the potential~\eq{eq:V} in which one of the fields was not self-interacting ($\lambda_1 \neq 0$ and $\lambda_2 = 0$). In our paper, we complement this work with the numerical analysis of the system in which both fields experience the self-interaction, $\lambda_{1,2} \neq 0$. We show that the simple extension (or, better to say, completion) of the model makes the analysis of the phase diagram very complicated. 

Let us start from the simplest case when the quartic interaction is absent: $\lambda_1 = \lambda_2 = 0$. The Hermitian~\eq{eq:mass:H} and non-Hermitian~\eq{eq:mass:NH} mass-squared matrices have the following eigenvalues, respectively:
\beqs
\beqn
M_{\HH,\pm}^2 & = & \frac{1}{2} \left( m_1^2 + m_2^2 \pm \sqrt{\left( m_1^2 - m_2^2\right)^2 + 4 m_5^4} \right), \qquad \\
M_{\NH,\pm}^2 & = & \frac{1}{2} \left( m_1^2 + m_2^2 \pm \sqrt{\left( m_1^2 - m_2^2\right)^2 - 4 m_5^4} \right). 
\eeqn
\label{eq:eigenmasses}
\eeqs
The vaccua in these models are stable provided the eigenmasses have no imaginary parts. For the Hermitian model with the mass matrix~\eq{eq:mass:H}, this requirement implies $M_{\HH,-}^2 \geqslant 0$ or, naturally, 
\beqn
\mbox{H}: \quad
\left\{
\begin{array}{c}
m_1^2 + m_2^2 > 0, \\[1mm]
m_1^2 m_2^2 - m_5^4\geqslant 0. 
\end{array}
\right.
\label{eq:stability:H}
\eeqn
These equations determine the region in the parameter space were the instability due to a negative mode (or, modes) at the trivial minimum, $\phi_1 = \phi_2 = 0$, does not occur. In an interacting theory with $\lambda_{1,2} \neq 0$, this instability corresponds to the spontaneous symmetry breaking. 

In the non-Hermitian model~\eq{eq:mass:NH}, the spectrum of the free (i.e., with $\lambda_1 = \lambda_2 = 0$) theory does not contain complex energy eigenvalues provided 
\beqn
\mbox{NH}: \quad 
\left\{
\begin{array}{c}
m_1^2 + m_2^2 > 0, \\[1mm]
m_1^2 m_2^2 + m_5^4 \geqslant 0, \\[1mm]
\left( m_1^2 - m_2^2\right)^2 - 4 m_5^4 \geqslant 0,
\end{array}
\right.
\label{eq:stability:NH}
\eeqn
where we also took into account a possibility that the squared masses $m_1^2$ and $m_2^2$ can take negative values. The first two requirements in Eq.~\eq{eq:stability:NH} correspond to the instability related to the spontaneous symmetry breaking rather than to the non-Hermiticity of the model. The last condition in Eq.~\eq{eq:stability:NH} highlights the region of the parameter space where the $\PT$ symmetry is said to be unbroken~\cite{Alexandre:2017foi,Mannheim:2018dur,Fring:2019hue}. Together, these conditions guarantee the stability of the ground state.

The classical equations of motion of the model~\eq{eq:L} are obtained by the variation of the Lagrangian~\eq{eq:L:prime} with respect to the independent fields $\phi_1^*$ and  $\phi_2^*$, respectively:
\beqs
\beqn
\square\phi_1+m_1^2\phi_1+m_5^2\phi_2+\frac{\partial V}{\partial\phi_1^*} & = & 0\,, \\
\square\phi_2+m_2^2\phi_2-m_5^2\phi_1+\frac{\partial V}{\partial\phi_2^*} & = & 0\,.
\eeqn
\label{eq:eqns:motion}
\eeqs
A variation with respect to the fields $\phi_1$ and $\phi_2$ gives us an inequivalent set of equations:
\beqs
\beqn
\square\phi_1^* + m_1^2\phi_1^* - m_5^2\phi_2^* + \frac{\partial V}{\partial\phi_1} & = & 0\,, \\
\square\phi_2^* + m_2^2\phi_2^* + m_5^2\phi_1^* + \frac{\partial V}{\partial\phi_2} & = & 0\,,
\eeqn
\label{eq:eqns:motion:star}
\eeqs
which differs from Eq.~\eq{eq:eqns:motion} by the sign flip of the off-diagonal mass term, $m_5^2 \to - m_5^2$.

The striking inconsistency of the two pairs of equations of motion, \eq{eq:eqns:motion} and \eq{eq:eqns:motion:star}, poses the natural question of how to treat this non-Hermitian system at the classical level. One of the earlier proposals~\cite{Alexandre:2017foi} suggests to use only one set of equations of motion, either \eq{eq:eqns:motion} or \eq{eq:eqns:motion:star}, and omit the complementary set. Indeed, the choice of the set flips the sign of the off-diagonal mass, $m_5^2 \to - m_5^2$, which does not affect the physical spectrum neither in free model~\eq{eq:eigenmasses} nor in the interacting model, as we will see below. Moreover, the non-Hermitian system must be open via coupling to an external source which equilibrates non-vanishing surface terms in the variation of the action.  When the off-diagonal mass vanishes, the system becomes Hermitian, and the single set of equations is enough to describe the solutions consistently.

\add{The next proposal was} to resolve the apparent problem with the inconsistency of the classical equations of motion was put forward in Ref.~\cite{Mannheim:2018dur} where a similarity transformation for the action has been used to achieve harmony between the two sets of equations of motion. This elegant idea initially required an extension of the field space using two complex components (with four degrees of freedom) for every original complex field (which hosts two degrees of freedom). 

The same mapping strategy has later been adapted and extended to many-field theories in Ref.~\cite{Fring:2019hue} where two real-valued fields were used to represent one complex field. However, this procedure, used for the mapping of the non-Hermitian theory to the Hermitian one via the similarity transformation, leads to the appearance of a negative kinetic energy term for one of the fields. While it was argued that this artifact does not change the signature of the appropriate Hilbert space~\cite{Mannheim:2018dur}, the appearance of the negative kinetic action leads to a negative contribution to the energy-momentum tensor so that the energy of a classical configuration (which is not necessarily a classical solution) would become unbounded from below. This pseudo-Hermitian method can be adapted, in specifying the appropriate physical regions, to give the classical solutions for non-Abelian monopoles in a spontaneously broken theory~\cite{Fring:2020xpi} (we refer the Reader to Refs.~\cite{Alexandre:2019jdb,Fring:2020bvr} for complementary discussions of a non-Hermitian non-Abelian gauge theory).

In our paper we follow the approach of Ref.~\cite{Alexandre:2017foi} where only one set of equations of motion -- either \eq{eq:eqns:motion} or \eq{eq:eqns:motion:star}, but not the both of them -- is considered. 
\add{
While this approach may seem restrictive, it still gives the complete description of classical and quantized theories~\cite{Alexandre:2020gah}. In addition to the invariance of the physical solutions with respect to the flip $m_5^2 \to - m_5^2$, the choice of the equations of motion coincides with the choice of the Hamiltonian operator in the quantum theory since both ${\hat H}(m_5^2)$ and ${\hat H}^\dagger(m_5^2) = {\hat H}(- m_5^2)$ can be used to promote the time evolution of the system. Moreover, the choice does not affect the non-Hermitian physics: the $\pm m_5^2$ versions of the classical equations of motion give the physically equivalent classical solutions of the theory and, in parallel, both original and Hermitian-conjugated Hamiltonians determine the very same evolution of the quantum theory~\cite{Alexandre:2020gah}.
}

\add{
The classical non-Hermitian theory becomes self-consistent if one supplements the Hermitian conjugation with the subsequent $\PT$ symmetric operation. Then the complex conjugation along with the $\PT$ flip $m_5^2 \to - m_5^2$ leaves the classical equations intact~\cite{Alexandre:2020gah}. The new combined operation is also important at the quantum level since the Hermitian-conjugated conversion between the bra and ket states should now be supplemented by the additional $\PT$ invariance operation in order to maintain the orthonormality of the eigenvectors of the non-Hermitian Hamiltonian~\cite{Alexandre:2017foi}. Thus, the would-be apparent non-equivalence of the original \eq{eq:eqns:motion} and conjugated \eq{eq:eqns:motion:star} equations does not lead to an inconsistency of the non-Hermitian theory. It gives the opposite: both formulations are consistent with lath other and lead to the same result both at classical and quantum levels.
}

At a certain stage, we use a numerical method to find the classical solutions using the criteria of the energy minimization as a selection principle of the right, ``minimal energy'' solution among all other available solutions. In our approach, the energy of the classical configuration in the non-Hermitian theory is bounded from below so that the numerical approach is self-consistent in finding the correct configuration. If we would otherwise employ the pseudo-Hermitian procedure of mapping the non-Hermitian theory to the Hermitian one, then the classical energy becomes unbounded due to the negative sign in the kinetic terms, and the numerical procedure fails to converge to a reasonable solution.

In the ground state the condensates are coordinate-independent quantities, and Eqs.~\eq{eq:eqns:motion} reduce to the non-linear algebraic relations:
\beqs
\begin{eqnarray}
m_1^2\phi_1+m_5^2\phi_2+2\lambda_1 \phi_1^2 \phi_1^* & = & 0\,, \\
m_2^2\phi_2-m_5^2\phi_1+2\lambda_2 \phi_2^2\phi_2^* & = & 0\,.
\end{eqnarray}
\label{eq:class:1}
\eeqs
The use of the complementary set of equations~\eq{eq:eqns:motion:star} instead of Eqs.~\eq{eq:eqns:motion} would lead to an equivalent physical solutions. Indeed, the swap of equations leads to the sign flip in Eq.~\eq{eq:class:1} which corresponds to a simple swap of the fields $\phi_1$ and $\phi_2$.

It is convenient to represent the fields $\phi_{a}$ in the radial form $\phi_a = v_a e^{i\theta_a}$ for $a=1,2$. Equations~\eq{eq:class:1} can possess nontrivial solutions provided the phases $\theta_a$ satisfy one of the following relations:
\beqn
\theta_1 = \theta_2, 
\qquad\
\theta_1 = \theta_2 + \pi\,. 
\label{eq:thetas}
\eeqn
One can cover both these cases assuming $\theta_1 = \theta_2$ and, simultaneously, allowing the amplitudes $v_a$ to take both positive and negative values, $v_a \in \R$. The classical equations~\eq{eq:class:1} then determine the amplitudes:
\beqs
\beqn
m_1^2 v_1 + m_5^2 v_2 + 2\lambda_1 v_1^3 & = & 0\,, \\
m_2^2 v_2 - m_5^2 v_1 + 2\lambda_2 v_2^3 & = & 0\,.
\eeqn
\label{eq:class:2}
\eeqs

The canonical energy-momentum of the model~\eq{eq:L} can be calculated by endowing the model with the gravitational background $g_{\mu\nu}$, considering the variation of the theory action, $S = \int d^4 \cL$ with respect to the metric,
\beqn
T^{\mu\nu} = - \frac{2}{\sqrt{-g}}  \frac{\delta S}{\delta g_{\mu\nu}(x)}\,,
\label{eq:Tmunu}
\eeqn
and then setting the metric to its Minkowski values back again, $g_{\mu\nu} \to \eta_{\mu\nu} \equiv \mathrm{diag}(+1,-1,-1,-1)$.
The energy density of the model~\eq{eq:L:prime} is given by the $\mu=\nu=0$ component of the energy-momentum tensor~\eq{eq:Tmunu}:
\beqn
T^{00} & = & \partial_0 \phi_1^* \partial_0 \phi_1 + {\bs \nabla} \phi_1^*  {\bs \nabla} \phi_1 
+ \partial_0 \phi_2^* \partial_0 \phi_2 + {\bs \nabla} \phi_2^*  {\bs \nabla} \phi_2 \nonumber\\
& & + m_1^2|\phi_1|^2 + m_2^2|\phi_2|^2 + m_5^2(\phi_1^*\phi_2 - \phi_2^*\phi_1) \nonumber \\
& & + \lambda_1 |\phi_1|^4 + \lambda_2 |\phi_2|^4\,.
\label{eq:T00}
\eeqn
One gets the following simple expression for the energy density of a uniform time-independent (ground) state:
\begin{equation}
E_{\NH,0} \equiv T^{00} = m_1^2 v_1^2 + m_2^2v_2^2 + \lambda_1 v_1^4 + \lambda_2 \phi_2^4\,.
\label{eq:E:NH}
\end{equation}
Both solutions~\eq{eq:thetas} for the phases $\theta_a \equiv \arg \phi_a$ of the condensates $\phi_a$ lead to the vanishing non-Hermitian term $\delta \cL_{\NH} = - 2 i m_5^2 \, {\mathrm{Im}}\,(\phi_1^*\phi_2) \equiv - 2 i m_5^2 v_1 v_2 \sin(\theta_1 - \theta_2)$ in the energy density~\eq{eq:E:NH}. Thus, the inconvenient imaginary terms do not enter the ground state of the model (we will see below that any solution of the single set of the classical equations of motion~\eq{eq:eqns:motion} has a real-valued energy density). However, despite the non-Hermitian mass $m_5$ not explicitly appearing in the vacuum energy~\eq{eq:E:NH}, it affects the ground state indirectly via the solutions of the equations of motion~\eq{eq:class:2}.

\begin{figure*}[htb]
\begin{tabular}{l|c|c|c}
& energy density & condensate 1 &  condensate 2 \\[2mm]
\hline
&&& \\[-2mm]
\rotatebox{90}{NH: $\quad m_1^2 < 0, \quad m_2^2 = |m_1|^2$} & 
\includegraphics[scale=0.25]{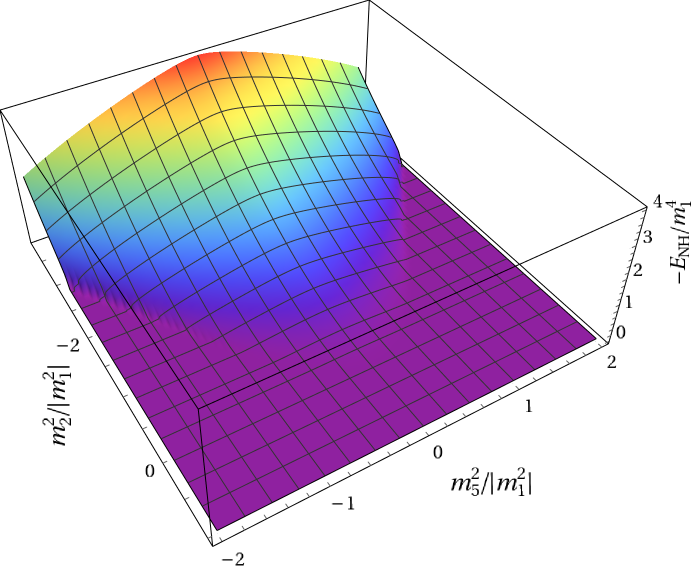} & 
\includegraphics[scale=0.25]{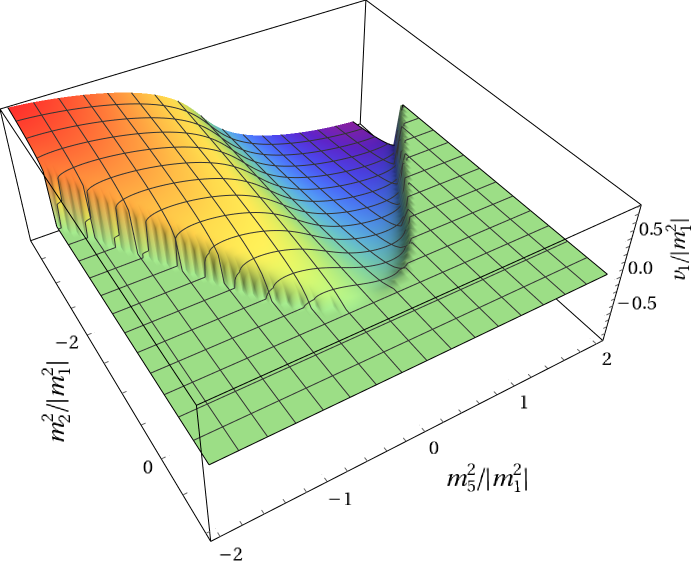} & 
\includegraphics[scale=0.25]{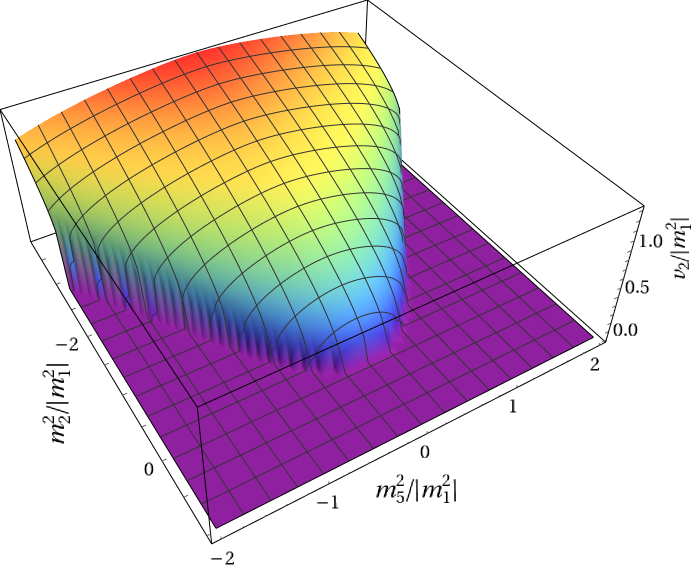} \\[-3mm]
& (a) & (b) & (c) \\[2mm]
\hline
& & & \\[-2mm]
\rotatebox{90}{NH: $\quad m_1^2 > 0, \quad m_2^2 = |m_1|^2$} & 
\includegraphics[scale=0.25]{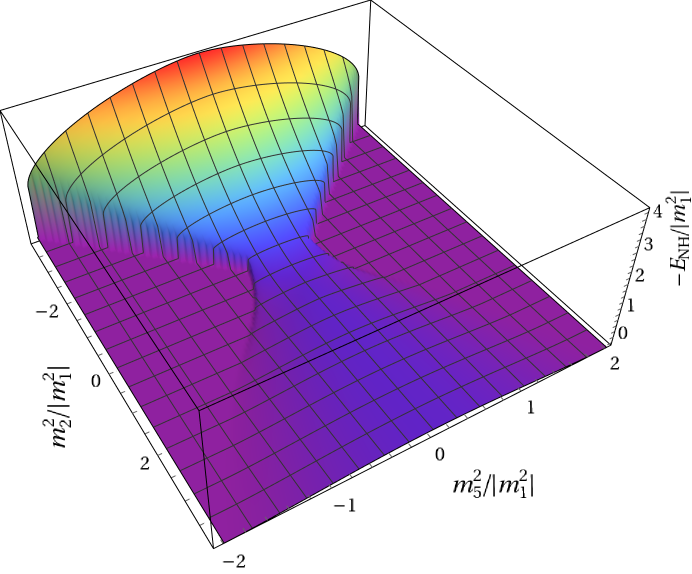} 
& 
\includegraphics[scale=0.25]{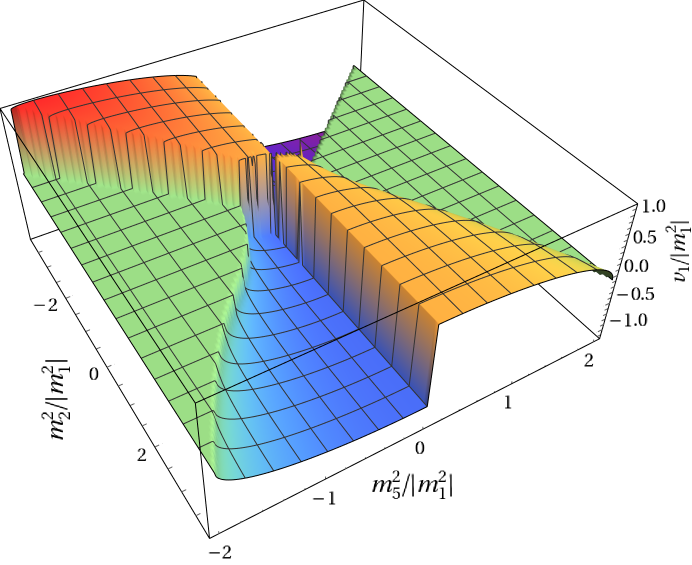} 
& 
\includegraphics[scale=0.25]{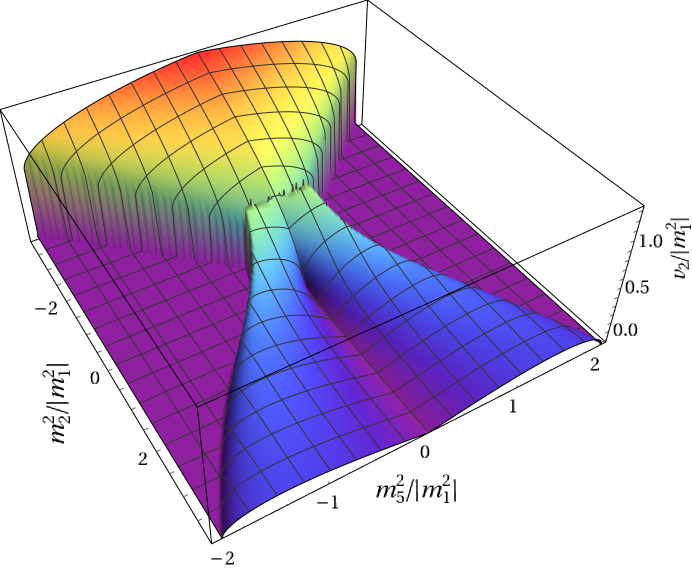} 
\\
& (d) & (e) & (f) \\[2mm]
\hline
& & & \\[-2mm]
\rotatebox{90}{H: $\quad m_1^2 < 0, \quad m_2^2 = |m_1|^2$} & 
\includegraphics[scale=0.25]{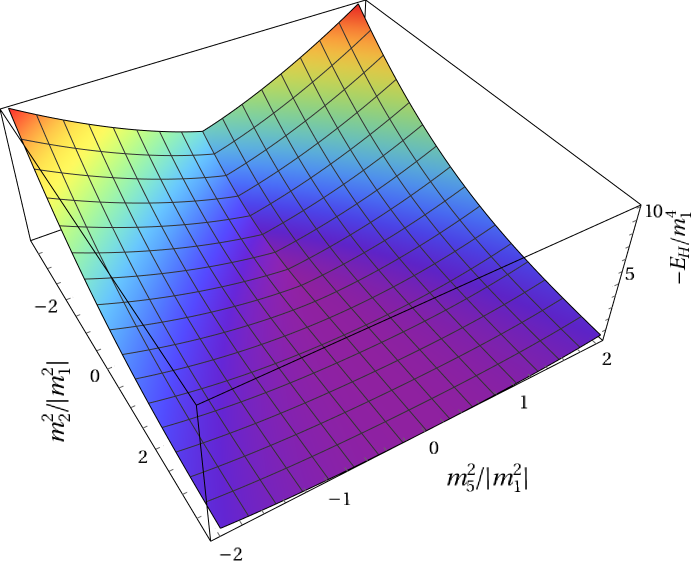} & 
\includegraphics[scale=0.25]{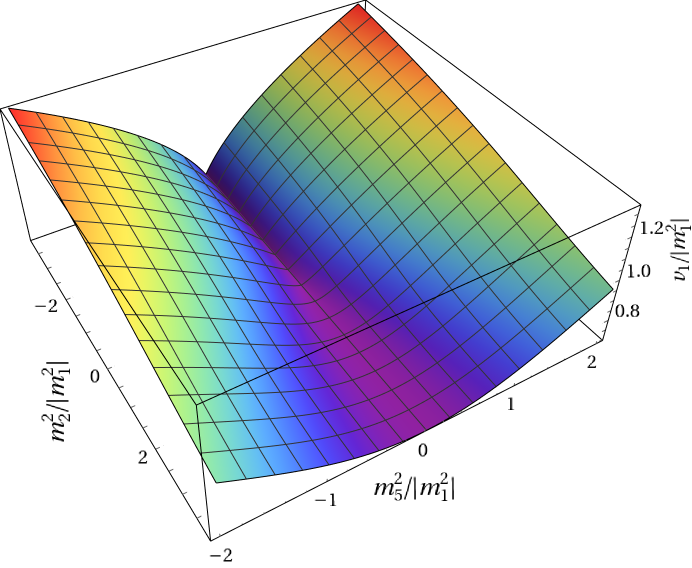} & 
\includegraphics[scale=0.25]{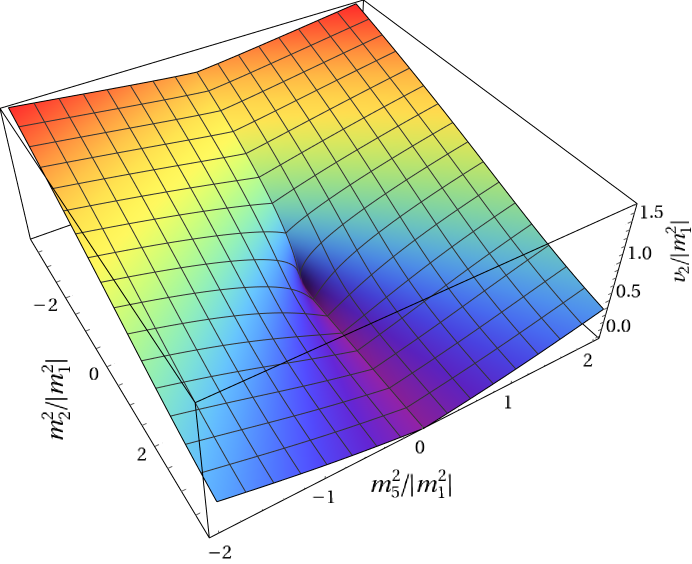} \\
& (g) & (h) & (i) 
\end{tabular}
\vskip 5mm
\caption{The upper panel: (a) the (minus) energy density~\eq{eq:E:NH} and the condensates (b) $v_1$ and (c) $v_2$ in the ground state of the non-Hermitian model~\eq{eq:L:prime} are shown in the plane of the mass parameter squared $m_2^2$ and the non-Hermitian mass squared $m_5^2$. The mass (squared) of the first field $\phi_1$ is taken positive, $m_1^2 > 0$, and the quartic couplings for both scalar fields are fixed: $\lambda_1 = \lambda_2 = 1$. The middle panel, with the plots (d), (e) and (f), corresponds to the same quantities obtained for the negative diagonal mass $m_1^2 < 0$. The lower panel, with the plots (g), (h) and (i), depicts the Hermitian case~\eq{eq:L:H}. All dimensionful quantities are shown in the units of the absolute value of the first mass parameter, $|m_1|$.}
\label{fig:energy:condensates}
\end{figure*}

\subsubsection{Hermitian ground state}
\label{sec:Hermitian:ground:state}

The equations of motion of the Hermitian model~\eq{eq:L:H} take the following form:
\beqs
\begin{eqnarray}
\square\phi_1+m_1^2\phi_1+m_5^2\phi_2+\frac{\partial V}{\partial\phi_1^*}=0\,,\\
\square\phi_2+m_2^2\phi_2+m_5^2\phi_1+\frac{\partial V}{\partial\phi_2^*}=0\,,
\end{eqnarray}
\label{eq:eqns:H}
\eeqs
which differ from Eq.~\eq{eq:eqns:motion} of the non-Hermitian model~\eq{eq:L:prime} as the both signs in front of the off-diagonal mass terms $m_5^2$ are the same in the Hermitian case~\eq{eq:eqns:H}. This small difference naturally propagates into the equations for the amplitudes in the ground state,
\beqs
\beqn
m_1^2 v_1 + m_5^2 v_2 + 2\lambda_1 v_1^3 & = & 0\,, \\
m_2^2 v_2 + m_5^2 v_1 + 2\lambda_2 v_2^3 & = & 0\,,
\eeqn
\label{eq:class:3}
\eeqs
as compared to its non-Hermitian analogue~\eq{eq:class:2}. However, in the striking dissimilarity with non-Hermitian case, the variations over the fields and their complex conjugates are obviously consistent with each other in the Hermitian model. Moreover, the energy density of the ground state of the Hermitian model,
\begin{equation}
E_{\HH,0}= m_1^2v_1^2 + m_2^2v_2^2 + 2 m_5^2 v_1 v_2 + \lambda_1 v_1^4+\lambda_2 v_2^4\,,
\label{eq:E:H}
\end{equation}
explicitly includes the $m_5^2$ coupling between the amplitudes of the different condensates. We remind that this coupling does not enter the non-Hermitian energy density~\eq{eq:E:NH}.

\subsubsection{Hermitian model vs. non-Hermitian model}

In the Hermitian model, a negative diagonal mass can trigger a natural spontaneous symmetry breaking of the global U(1) symmetry~\eq{eq:U1} which leads to a non-zero expectation value of the doublet field, $\Phi_0 \neq 0$. The runaway of the scalar field from the symmetric state is balanced by quartic (self)-interactions so that the symmetry-broken ground state of the system is stable. In an interacting model, the stability of the symmetric, $\Phi = 0$, state is determined by Eq.~\eq{eq:stability:H}.

The interacting non-Hermitian theory possesses two types of instabilities. In addition to the mentioned spontaneous symmetry breaking stipulated by the first two relations in Eq.~\eq{eq:stability:NH}, the symmetric ground state can also experience the U(1) symmetry breaking which could be caused, in turn, by the broken discrete $\PT$ symmetry. This purely non-Hermitian effect is dictated by the third relation in Eq.~\eq{eq:stability:NH}, the violation of which leads to the complex mass spectrum and, consequently, to the instability. The $\PT$-induced instability could be well seen when both diagonal masses are positive so that the first two relations in Eq.~\eq{eq:stability:NH} are satisfied. As in the purely Hermitian case, the symmetry-broken ground state of the non-Hermitian system is expected to be stabilized by the quartic (self)-interactions.

In Fig.~\ref{fig:energy:condensates} we compare the U(1) symmetry breaking patterns in the non-Hermitian model with $m_1^2>0$ (the upper panel) and  $m_1^2<0$ (the middle panel) as well as the Hermitian model with $m_1^2 <0$ (the lower panel). We show the total energy density $E$ determined by Eqs.~\eq{eq:E:NH} and \eq{eq:E:H}, and the condensates $v_1$ and $v_2$. Notice that the non-Hermitian~\eq{eq:class:2} and Hermitian~\eq{eq:class:3} equations on the ground state imply that the relative sign of the condensates is determined by the ground state while the overall sign is not fixed. To remove this arbitrariness, we require the condensate $v_2$ to be positive. 

If both diagonal squared masses are positive, $m_1^2 >0$ and $m_2^2 >0$, and the off-diagonal mass is zero, $m_5 = 0$, then the ground state resides in the symmetric phase with vanishing condensates. As the off-diagonal mass increases in its absolute value, the theory should experience the $\PT$-symmetric instability and we expect the appearance of the condensates in the $(m_1^2>0,m_2^2>0)$ plane. However, this does not happen, as one can see in Figs.~\eq{fig:energy:condensates}(b) and (c). The presence of the ``$\PT$-induced'' condensates is forbidden by the requirement of the minimization of the energy density~\eq{eq:E:NH} as these condensates would have made the energy density positive while the trivial ground state with $v_1 = v_2 = 0$ has zero energy. This conclusion appears to be the consequence of the fact that the expression for the non-Hermitian energy density~\eq{eq:E:NH} does not contain a ``compensating'' $m_5^2$ term which is present, on the contrary, in the expression of the Hermitian energy density~\eq{eq:E:H}. Therefore, the U(1) symmetric ground state with vanishing energy is preferred over the U(1) broken state with the nonvanishing condensates. We will see below that some of these symmetric states belong to the $\PT$-broken phases of the interacting model and, therefore, are physically meaningless. 

In the symmetry-broken phase with $m_2^2 < 0$ (while we still keep $m_1^2 > 0$), the conventional U(1) symmetry breaking does occur. The upper panel of Fig.~\ref{fig:energy:condensates} shows that the non-Hermitian mass influences the conventional symmetry breaking in a somewhat controversial way: at sufficiently large $m_5^4$, the third requirement of Eq.~\eq{eq:stability:NH} is violated, the $\PT$ symmetry gets broken and the U(1) symmetry gets restored because the ground-state condensates $v_1$ and $v_2$ vanish. In the symmetry-broken region, the flip of the sign in the $m_5^2$ mass flips the sign of the $v_1$ condensate (we remind that we always keep the $v_2$ condensate non-negative). 

A rather similar picture occurs when one of the diagonal mass squared is taken to be negative, $m_1^2 < 0$, as shown in the middle panel of Fig.~\ref{fig:energy:condensates}. The spontaneous symmetry breaking occurs at both signs of the remaining mass squared, $m_2^2$, while the increase of the absolute value of the off-diagonal mass $m_5^2$ leads to the restoration of the U(1) symmetry. We would like to stress again that this particular picture is enforced by the requirement of minimization of the non-Hermitian energy density~\eq{eq:E:NH} which allows us to choose the correct ground state from the multitude of the solutions of the non-Hermitian equations of motion~\eq{eq:class:2}. 

Finally, the ground state of the Hermitian two-scalar model with $m_1^2 < 0$ is shown in the lower panel of Fig.~\ref{fig:energy:condensates}. The condensates appear almost at every point of the phase diagram except for the line $m_5=0$, where the condensates decouple and the $v_2$ condensate ceases to exist at $m_2^2>0$. At this semi-infinite line, the condensate $v_1$ takes its minimum. 

\subsubsection{Stability of the ground state and the $\PT$ symmetry in the non-Hermitian model}
\label{sec:nonHermitian:ground:state}

Before proceeding to the discussion of the vortex solutions, let us address the formal stability issues of the ground state. Usually, the local stability of a classical configuration is probed by expanding the scalar fields in the vicinity of the configuration, $\phi_{a}=v_{a} + {\hat\phi}_{a}$ with $|{\hat\phi}_{a}| \ll |v_a|$. The configuration is unstable if the fluctuation matrix corresponding to the variation of the action with respect to the fluctuation of the fields contains negative modes.

In the Hermitian theory, the global minimum in the total energy of the solution corresponds to an absolutely stable state. In the non-Hermitian theory, this criterion may not work since even in the classical theory, the ground state is determined by a single set of classical equations of motion out of the existing two sets. Therefore, mathematically, one could expect the emergence of negative directions in the space of fields. This expectation is coherent with our physical intuition since the non-Hermitian system resides in the steady state which is generally not a thermodynamic equilibrium. In the remaining part of this section, we argue that the formal stability criteria can still be applied to the interacting non-Hermitian systems. These criteria give us the generalization of the $\PT$-symmetric and $\PT$-broken regions for the interacting theory.

From the classical equations of motion~\eq{eq:class:1}, we obtain that the field fluctuations around the condensates are governed by the following equation:
\begin{equation}
\square
\begin{pmatrix}
\hat\phi_1\\
\hat\phi_1^*\\
\hat\phi_2\\
\hat\phi_2^*
\end{pmatrix}
+ {\mathcal M}^2
\begin{pmatrix}
\hat\phi_1\\
\hat\phi_1^*\\
\hat\phi_2\\
\hat\phi_2^*
\end{pmatrix}=0,
\label{eq:F:M}
\end{equation}
where the fluctuation matrix ${\mathcal M}^2_\NH$ of the non-Hermitian model is as follows:
\beqn
{\mathcal M}^2_\NH {=}
\begin{pmatrix}
4 \lambda _1 v_1^2{+}m_1^2 & 2 v_1^2 \lambda _1 & m_5^2 & 0 \\
 2 v_1^2 \lambda _1 & 4 \lambda _1 v_1^2{+}m_1^2 & 0 & m_5^2 \\
 -m_5^2 & 0 & 4 \lambda _2 v_2^2{+}m_2^2 & 2 v_2^2 \lambda _2 \\
 0 & -m_5^2 & 2 v_2^2 \lambda _2 & 4 \lambda _2 v_2^2{+}m_2^2 \\
\end{pmatrix}.\
\nonumber
\eeqn
In the U(1) broken phase, this matrix has one zero eigenvalue which corresponds to the Goldstone mode. In the symmetric U(1) phase, all eigenvalues are generally nonzero. 

\begin{figure*}
\begin{tabular}{c|c|c|c|c}
& \hskip 3mm $\lambda_2=0$ & $\lambda_2=0.1$ & \hskip 3mm $\lambda_2=1$ & $\lambda_2=2$ \\
\hline 
& & & & \\[-2mm]
\rotatebox{90}{\hskip 12mm $\quad m_1^2 > 0$} & 
\includegraphics[scale=0.2]{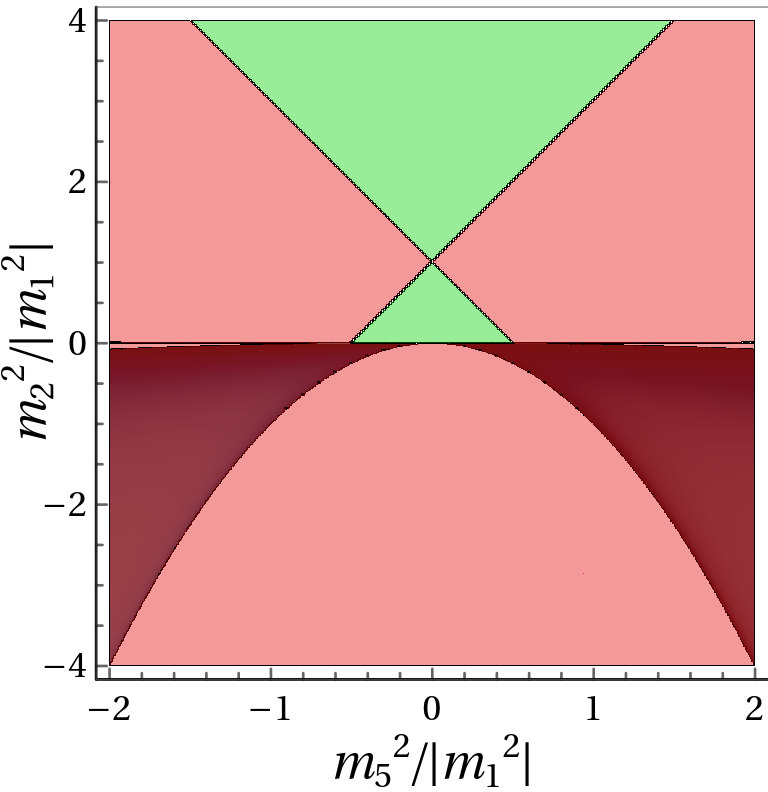} & \hskip 3mm 
\includegraphics[scale=0.2]{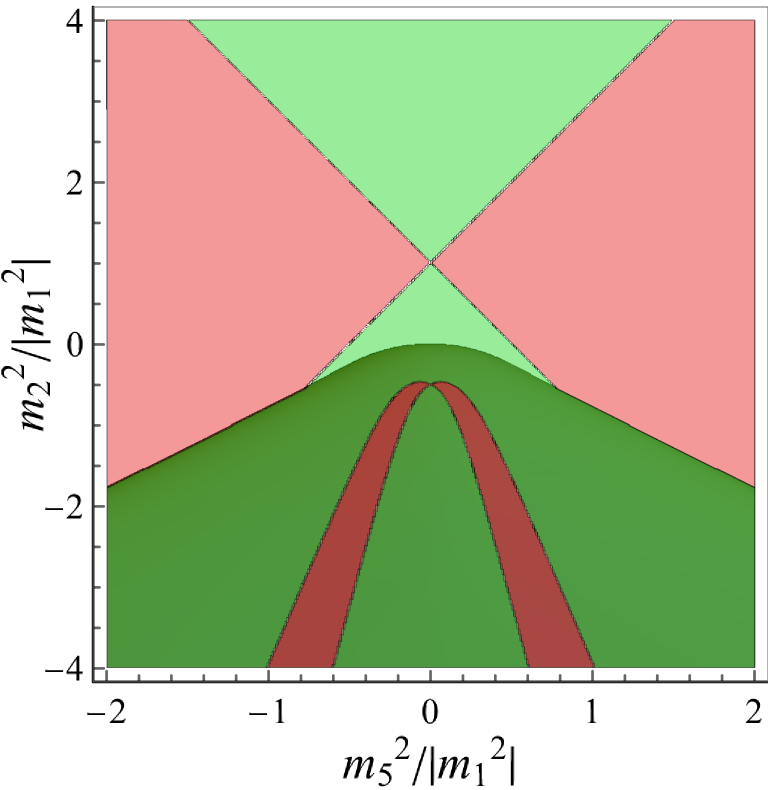} & \hskip 3mm 
\includegraphics[scale=0.2]{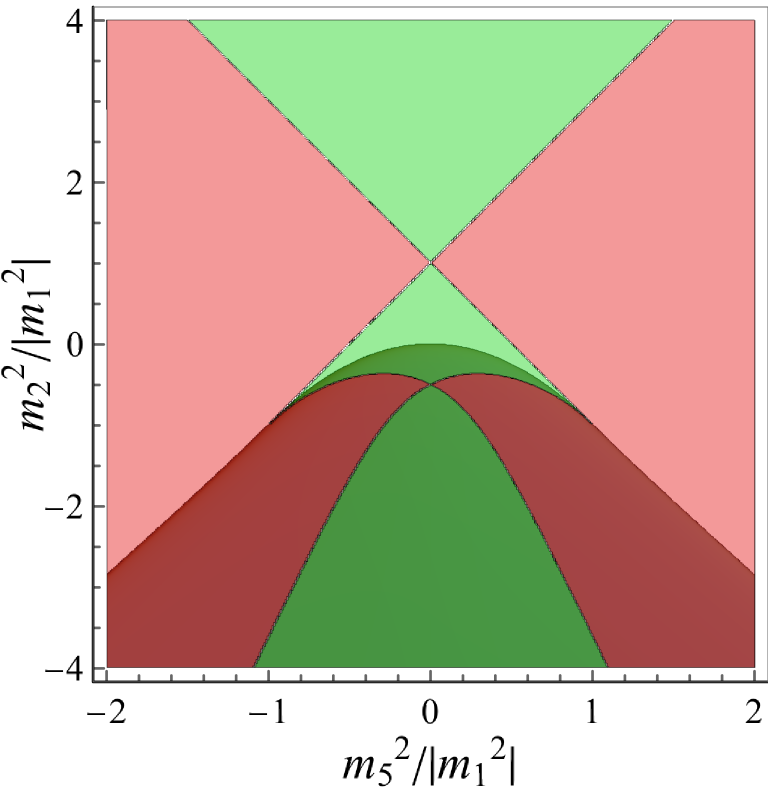} &  \hskip 3mm 
\includegraphics[scale=0.2]{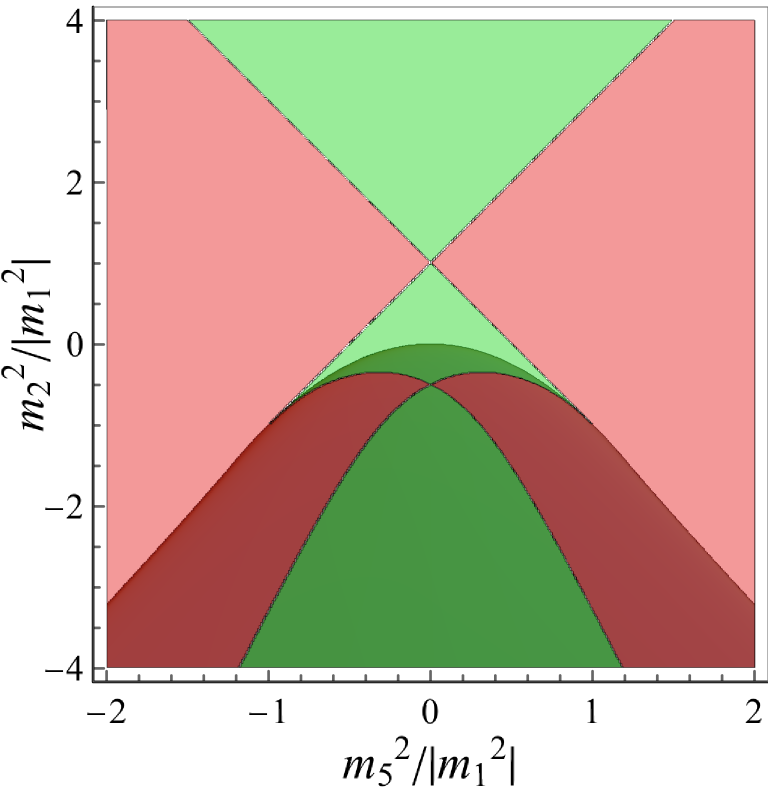} \\
& (a) & (b) & (c) & (d) \\[1mm]
\hline
& & & & \\[-2mm]
\rotatebox{90}{\hskip 12mm $\quad m_1^2 < 0$} & 
\includegraphics[scale=0.2]{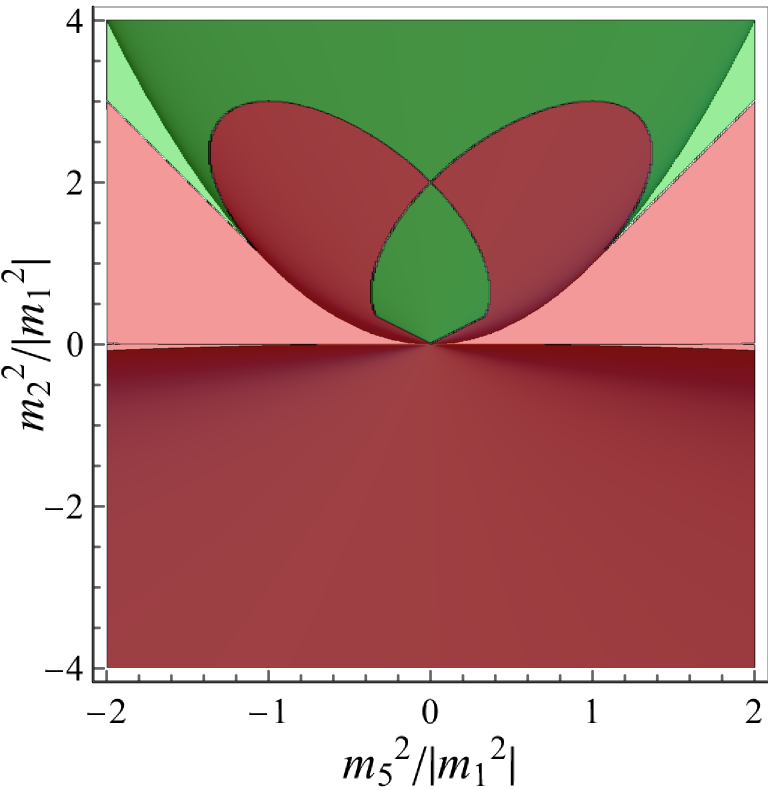} & 
\includegraphics[scale=0.2]{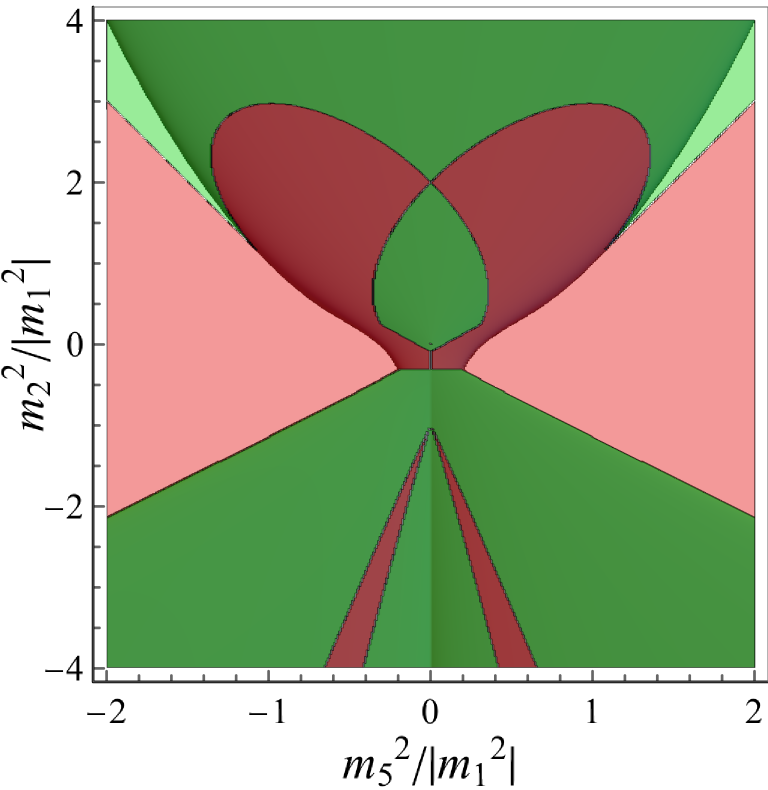} & 
\includegraphics[scale=0.2]{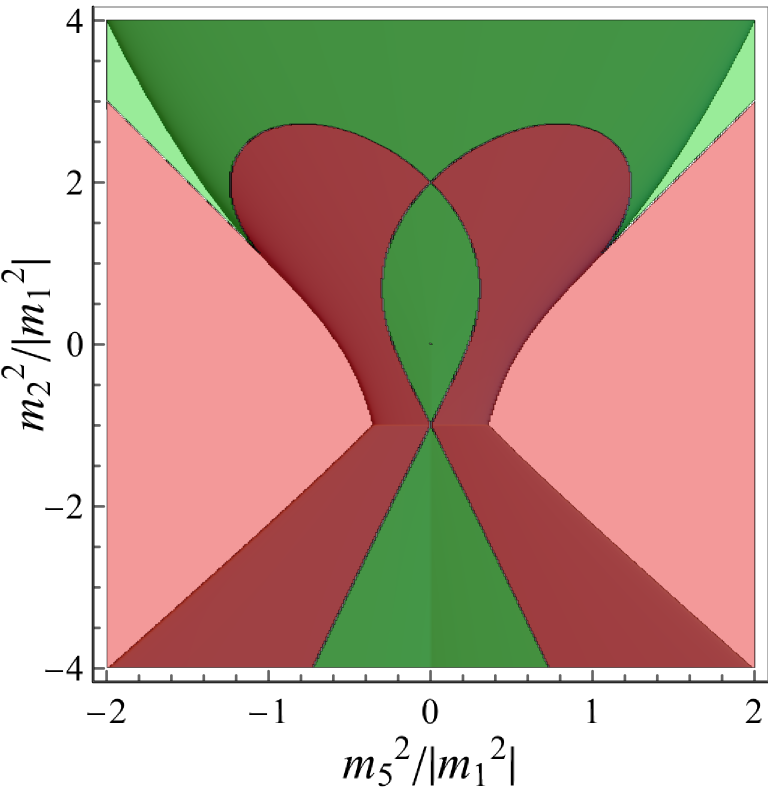} &  
\includegraphics[scale=0.2]{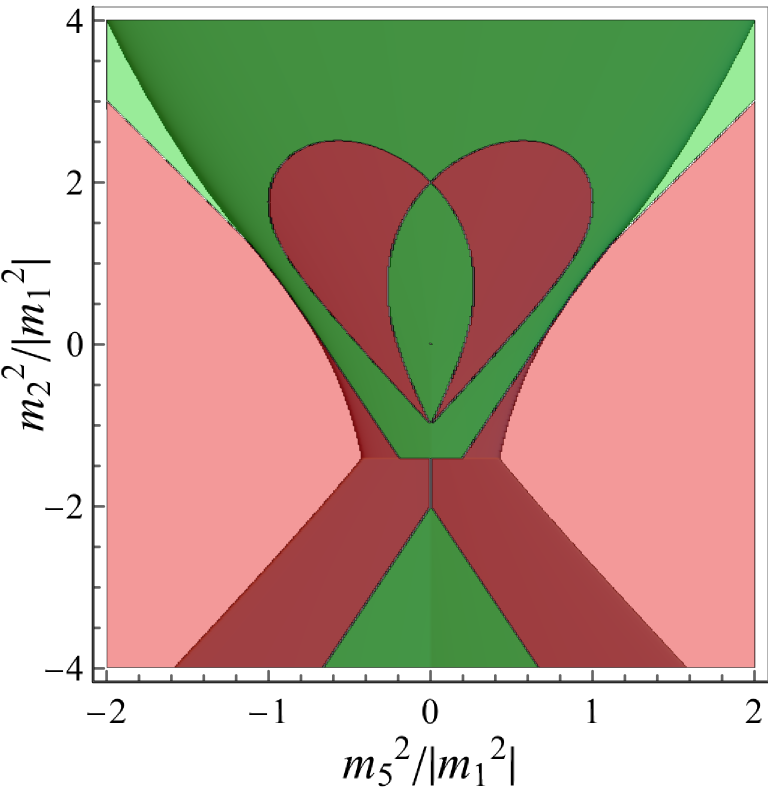} \\
& (e) & (f) & (g) & (h)
\end{tabular}
\caption{The stable ($\PT$-symmetric, marked by the green color) region and the unstable (spontaneously $\PT$-broken, marked by the red color) regions in the interacting non-Hermitian theory in the $(m_5^2, m_2^2)$ plane. The mass squared of the first scalar field takes positive values ($m_1^2 > 0$) at the upper panel and negative values ($m_1^2 < 0$) at the lower panel. The dark green and light green colors denote regions with stable $\PT$-symmetric ground states with non-vanishing and vanishing condensates, respectively. The dark red and light red areas are unstable $\PT$-broken regions with, respectively, non-trivial and vanishing solutions of the classical equations of motion~\eq{eq:eqns:H}.
}
\label{fig:stability:areas:NH}
\end{figure*}

For the ground state to be stable, one expects that the all eigenvalues of the fluctuation matrix are non-negative. In the non-Hermitian model, this requirement is not always satisfied. In Fig.~\ref{fig:stability:areas:NH} we show the (in)stability phases for various quartic couplings $\lambda_1$ and $\lambda_2$. 

The (in)stability phase diagram has a rather curious form. One notices that the borderlines between the stable and unstable areas involve both U(1) symmetric and broken regions which possess, on the contrary, rather featureless, smooth behavior of the condensates as shown in Fig.~\ref{fig:energy:condensates}. In order to understand the appearance of the negative modes, let us consider any unstable point in the symmetry-restored region with masses $m_1^2>0$, $m_2^2>0$ (these regions are marked by the white color in the upper panel of Fig.~\ref{fig:stability:areas:NH}). Since the both condensates are zero and both diagonal masses as well as the couplings are positive, the energy density~\eq{eq:E:NH} takes an absolute minimum. On the other hand, in these regions, the third criterion of Eq.~\eq{eq:stability:NH} is not satisfied, indicating that the symmetric state resides in the $\PT$-broken region and is thus unstable. The instability is not captured by the minimization of the energy density~\eq{eq:E:NH} since this particular expression is derived for the classical solutions while the instability can drive the configuration out of the classical subspace of configurations. The non-classical configuration can acquire even a complex value thus invalidating the criterion of the energy minimization outside of the classical subspace of field configurations. 

Thus, we arrive to the conclusion that the ground state of the non-Hermitian model is formally not stable in certain regions of the parameter space. One can easily check that these unstable regions become the $\PT$-broken regions when the quartic interaction couplings are set to zero, $\lambda_{1,2} \to 0$. Therefore, similarly to the free non-interacting model, the unstable regions indicated in Fig.~\ref{fig:stability:areas:NH} can be interpreted as the $\PT$-broken regions in the interacting case where the model cannot be used as a invalid prescription of any steady state in a physical system. The $\PT$-symmetric regions, on the contrary, are valid stable zones where the steady-state physics can be realized. 

The stability, now associated with the $\PT$-symmetric regions, is thus determined by the standard requirement that the quadratic fluctuation matrix, 
\beqn
{\mathcal M}^2_{ab} = \frac{\delta^2 S_\NH [\chi]}{\delta \chi_a(x) \delta \chi_b(x)}\,,
\eeqn
does not possess negative eigenvalues. Here the vector ${\vec \chi} = (\phi_1, \phi_2, \phi_1^*, \phi_2^*)$ denotes the original fields and their conjugates.

For the sake of completeness, we show the fluctuation matrix of the Hermitian model~\eq{eq:L:H},
\beqn
{\mathcal M}^2_\HH {=} 
\begin{pmatrix}
4 \lambda _1 v_1^2{+}m_1^2 & 2 v_1^2 \lambda_1 & m_5^2 & 0 \\
 2 v_1^2 \lambda _1 & 4 \lambda_1 v_1^2+m_1^2 & 0 & m_5^2 \\
 m_5^2 & 0 & 4 \lambda_2 v_2^2 {+} m_2^2 & 2 v_2^2 \lambda _2 \\
 0 & m_5^2 & 2 v_2^2 \lambda _2 & 4 \lambda _2 v_2^2 {+} m_2^2 \\
\end{pmatrix},
\nonumber
\eeqn
which has only non-negative eigenmodes, as expected. The Hermitian ground state does not have any ambiguities in the stability criteria.

\section{Vortices in fixed-length condensates}
\label{sec:vortices:London}

\subsection{Vortices in one-component superfluids: brief review}

\add{In this section, we consider in the bosonic condensates with the fixed radial (amplitude) part $|\psi|$ of the scalar condensate $\psi = |\psi| e^{i \theta}$ when the entire dynamics of the system is encoded in the condensate phase $\theta$. We can neglect the fluctuations of the amplitude in the long-range limit for fluctuations with a wavelength longer than the appropriate healing length $\xi$~\cite{ref:Pitaevskii:book}.}

\add{
The healing length corresponds to the wavelength at which the kinetic energy of the boson equals the chemical potential.
It determines the size of vortices and plays a role similar to the one played by the correlation length in the superconductivity.} On the superconductivity side, a condensate with a fixed amplitude appears naturally in the London limit which corresponds to the extreme type II superconducting regime. In the London limit, the penetration depth of the magnetic field is much longer than the correlation length so that the latter can be neglected in many physical situations. Equivalently, this limit poses a constraint on the modulus of the superconducting condensate which can be considered as a rigidly fixed, non-fluctuating quantity. The gauge invariance implies that the phase of the superconducting condensate is not constrained by this limit, thus sharing an analogy with a superfluid. However, the phase is absorbed by the electromagnetic gauge field, which becomes -- consistent with the Meissner effect -- a massive vector field via the Anderson-Higgs mechanism.~\cite{book:LL}

In the background of sufficiently strong magnetic field, the type-II superconductors enter the mixed (Abrikosov) phase where the magnetic flux penetrates the condensate in a form of parallel vortices which form the regular Abrikosov lattice. In the London limit, the vortex is characterized by a singularity in the phase of the condensate which reflects itself in a singular behavior of the gauge field close to the vortex core.~\cite{book:Kleinert}

In our paper, we consider the model with the global, rather than local/gauge, U(1) symmetry. In the global case, the condensate is associated with the electrically neutral complex scalar field which corresponds to the phenomenon of superfluidity rather than superconductivity.  The Anderson-Higgs mechanism is evidently absent and the massless mode in the phase of the condensate appears to be the Goldstone boson associated with the global U(1) symmetry breaking. In the case of a non-relativistic Bose gas, one identifies the Goldstone mode with a phonon. Instead of the Abrikosov strings with magnetic flux, the superfluid condensates host the vortices characterized by singularities in the phases of the condensate. These vortices are also called the ``global vortices'' since they appear in theories possessing a global symmetry. Below, we will briefly review the theory of vortices for a non-relativistic one-component superfluid following Ref.~\cite{book:Kleinert} and then we proceed to the generalization of our approach to the long-range (``London'') limit of the relativistic non-Hermitian two-field model. 

Consider a superfluid condensate of a non-relativistic Bose gas of particles with the mass $M$. It is convenient to describe by the wave function in the radial coordinates $\psi = |\psi| e^{i \theta}$. The long-range limit corresponds to the fixed radial degree of freedom $|\psi|$ which is a good approximation at very low temperature. Then the energy of the fluctuations in the superfluid can be written in the following form (we work in units $c = \hbar = 1$):
\beqn
E = \int d^3 x \, \frac{\rho_s}{2} {\bs v}_s^2 \equiv \frac{|\psi|^2}{2 M} \int d^3 x ({\bs \nabla} \theta)^2  \,,
\label{eq:E:non-rel}
\eeqn
where the radial part $|\psi|$ of the condensate determines the superfluid density $\rho_s = M |\psi|^2$ while its phase $\theta$ provides us with the velocity of the superfluid condensate, ${\bs v}_s = ({\bs \nabla} \theta)/M$. We neglected here qualitatively inessential contributions coming from the inhomogeneity of the radial part of the condensate. An extremization of the energy~\eq{eq:E:non-rel} gives us the Poisson's wave equation, $\Delta \theta = 0$ which corresponds to the incomprehensibility condition of the superfluid, ${\bs \nabla} \cdot {\bs v}_s = 0$. 

The superfluid theory~\eq{eq:E:non-rel} also incorporates singular configurations of the vortex fluid which correspond to the superfluid currents that wind around a line (called the vortex line) in three-dimensional space. Precisely at the vortex line, the superfluid condensate vanishes, $\phi = 0$, while the phase of the superfluid condensate is equal -- in the vicinity to the vortex core -- to a geometrical angle (times an integer) in the plane transverse to the vortex line. The integer is a topological quantity which characterizes the vortex winding number. 

The vortices can be accounted in the model~\eq{eq:E:non-rel} despite the fact that this model describes a long-range macroscopic physics with a globally uniform, constant condensate. Consider, for simplicity, a straight static vortex line along the direction $x_3$ located at the center $x_1 = x_2 = 0$ in the transverse plane. The phase of the vortex is $\theta(x_1,x_2) = n \varphi$ where $n$ is the winding number of the vortex and $\varphi = \arg(x_1 + i x_2)$ is the azimuthal angle in the two-dimensional $(x_1,x_2)$ plane. The velocity of the superfluid, 
\beqn
{\bs v}_s = \frac{n}{M} {\bs \nabla} \arg(x_1 + i x_2) = \frac{n}{M r} {\bf e}_\varphi,
\eeqn
is directed along the unit vector in the azimuthal ${\bf e}_\varphi$, so that the fluid ``winds'' around the vortex center at $x_1 = x_2 = 0$. Alternatively, the location of the vortex singularity corresponds to the point in the two-dimensional plane, where the derivatives do not commute, 
\beqn
(\partial_1 \partial_2 - \partial_2 \partial_1) \theta(x_1,x_2) = 2 \pi n \delta(x_1) \delta(x_2)\,.
\label{eq:noncommutativity:nr}
\eeqn

The total energy~\eq{eq:E:non-rel} evaluated at this static straight vortex configuration is:
\beqn
E = L \frac{|\psi|^2 n^2}{2 M} 2 \pi \int_\xi^R r d r \frac{1}{r^2} = L \frac{\pi |\psi|^2 n^2}{M} \ln \frac{R}{\xi}\,,
\label{eq:energy:vortex:nr}
\eeqn
where $L$ is the length of the vortex. The infrared cutoff of this integral is the size of the system $R$ while the size $\xi$ of the vortex core -- typically, of the interatomic distance -- serves as an ultraviolet cutoff. The important message of Eq.~\eq{eq:energy:vortex:nr} is that the vortex energy (mass) is proportional to the length of the vortex $L$. One can show that the vortex curvature provides a subleading correction. The energy density per unit length is a finite quantity since the logarithmic divergence in Eq.~\eq{eq:energy:vortex:nr} is very mild.

Consider now the relativistic one-component model with the action:
\beqn
S = \int d^4 x\, \partial_\mu \phi^* \partial^\mu \phi \equiv \kappa^2 \int d^4 x\, \partial_\mu \theta \partial^\mu \theta\,,
\label{eq:S:XY}
\eeqn
where we adopted the long-range limit with the constant radial condensate $\kappa = |\phi|$. The variation of the action with respect to the phase $\theta$ gives us the equation for the propagation of the massless Goldstone particle, $\Box \, \theta = 0$, where $\Box = \partial_t^2 - {\bs \nabla}^2$ is the d'Alembert operator.

In the non-relativistic limit, the model~\eq{eq:S:XY} possesses the energy~\eq{eq:E:non-rel} for static vortex configurations. It is convenient to parameterize the coordinates ${\tilde x}^\mu = {\tilde x}^\mu(\vec\sigma)$ of the two-dimensional vortex singularities by the two component vector ${\vec \sigma} = (\sigma_1, \sigma_2)$. We split the phase $\theta$ into the regular and singular parts, $\theta = \theta^r + \theta^s$. The regular part $\theta^r$ of the phase corresponds to the perturbative fluctuations while the singular part $\theta^s$ encodes the position of the vortex~\eq{eq:noncommutativity:nr}. In the relativistic notations,
\beqn
\partial_{[\mu,} \partial_{\nu]} \theta^s (x, \tilde x) = 2 \pi \cdot \frac{1}{2} \epsilon_{\mu\nu \alpha \beta} 
\Sigma_{\alpha \beta}(x, \tilde x), 
\label{eq:theta:sing}
\eeqn
where the singularity itself is given by the tensor current: 
\beqn
\Sigma_{\alpha \beta}(x, \tilde x) = \int_{\Sigma} d^2 
\sigma_{\alpha \beta}(\tilde x) \delta^{(4)} [x -\tilde x(\vec\sigma)],
\label{eq:Sigma}
\eeqn
which is expressed via the differential measure at the vortex world-sheet $\Sigma$:
\beqn
d^2 \sigma_{\alpha \beta}(\tilde x) = \epsilon^{ab} 
\frac{\partial x_{\alpha}}{\partial \sigma_a}  \frac{\partial x_{\beta}}{\partial \sigma_b} d^2 \sigma \,.
\eeqn
Here $\epsilon^{ab}$ is the fully anti-symmetrized tensor in two dimensions, $\epsilon^{12} = - \epsilon^{21} = + 1$ and $\epsilon^{11} = \epsilon^{22} = 0$. The vortex tensor~\eq{eq:Sigma} is a two-dimensional delta-function at the surface of the vortex, with the orientation at the vortex world-sheet. For example, for a straight vortex mentioned above, one uses the parameterization ${\tilde x}_0  = \sigma_1$,  ${\tilde x}_1 = {\tilde x}_2 = 0$, ${\tilde x}_3  = \sigma_2$, and obtains 
\beqn
\Sigma_{\alpha \beta}(x) = n \,
(\delta_{\alpha,0} \delta_{\beta,3} - \delta_{\alpha,3} \delta_{\beta,0}) \delta(x_1) \delta(x_2)\,,
\label{eq:example}
\eeqn 
where $n \in \Z$ is the vorticity. Note that the regular part of the phase does not contain any singularity by definition, $[\partial_{\mu}, \partial_{\nu}] \theta^r \equiv 0$.

Integrating out or, equivalently, solving the equations of motion for the regular component the phase, $\theta^r$, allows us to rewrite the action~\eq{eq:S:XY} in terms of the singular part of the phase $\theta^s$, which, in turn, depends only on the vortex world-sheet~\eq{eq:Sigma}:
\beqn
S[\Sigma] = 4 \pi^2 \kappa^2 \int_\Sigma d^2 \sigma(\tilde x) \int_\Sigma d^2 \sigma({\tilde x}') D({\tilde x} - {\tilde x}')\,.
\label{eq:S:Sigma}
\eeqn
This action is a nonlocal functional which features two integrals that are taken over the same vortex worldsheet. In the case of many vortices, the worldsheet $\Sigma$ includes all their worldsheets: $\Sigma_{\mu\nu} = \Sigma_{1,\mu\nu} + \Sigma_{2,\mu\nu} + \dots$.

The nonlocal action~\eq{eq:S:Sigma} represents the self-interaction of the vortex line as well as the interactions of the distinct vortex segments via propagation of a massless Goldstone particle between the vortex segments. In Eq.~\eq{eq:S:Sigma}, this long-range interaction is represented by the advanced Green's function $D(x)$ of the d'Alembert operator: 
\beqn
\Box \, D(x) = - \delta^{(4)}(x)\,.
\label{eq:D}
\eeqn

In the case of a static straight vortex line of large length $L$, the action~\eq{eq:S:Sigma} calculated for the time interval $\delta t$ gives us $S = E \delta t$, where $E$ is, up to parameter redefinitions, the known vortex energy~\eq{eq:energy:vortex:nr}. In order to demonstrate this fact, it is convenient to make a Wick transformation in the integral~\eq{eq:S:Sigma} to the Euclidean spacetime. In the Euclidean space, the massless propagator~\eq{eq:D} is:
\beqn
D(x) = \int \frac{d^4 p}{(2 \pi)^4} \frac{e^{i p x}}{p^2} = \frac{1}{8 \pi^2} \frac{1}{|x|^2}\,,
\eeqn
where $x$ is the 4-distance. For a single straight static vortex with the surface~\eq{eq:example}, the longitudinal (along the vortex) and temporal coordinates in the action~\eq{eq:S:Sigma} can be integrated out, and we get the following formal expression for the vortex energy:
\beqn
E =  4 \pi^2 \kappa^2 L \, D^{(2d)} (0)\,,
\label{eq:E:0}
\eeqn
where 
\beqn
D^{(2d)} (\rho) = - \frac{1}{2\pi} \ln \frac{\rho}{\rho_0}\,,
\eeqn
is the two-dimensional massless propagator [a solution of Eq.~\eq{eq:D} in two Euclidean dimensions] as the function of the two-dimensional distance $\rho$. The parameter $\rho_0$, which has the dimension ``length'', is introduced for the consistency reasons. The argument ``0'' in Eq.~\eq{eq:E:0} highlights the fact that the formal expression for the energy in the long-range limit is a logarithmically divergent quantity similarly to the non-relativistic expression~\eq{eq:energy:vortex:nr}. A more accurate derivation in a finite cylindrical box of the radius $R_0$ leads us to
\beqn
E =  2 \pi \kappa^2 L \log \frac{R_0}{\xi}\,,
\label{eq:E:1}
\eeqn
where $\xi$ is the size of the vortex core \add{given by the healing length which does not explicitly enter the model in the long-wavelength limit.}

The general expression~\eq{eq:S:Sigma} also gives us the interaction energy of the two straight static vortices with the vorticities $n_1$ and $n_2$ separated by the distance $R$:
\beqn
V(R) & = & 8 \pi^2 \kappa^2 L D^{(2d)} (R) \nonumber \\
& \equiv & - 4 \pi n_1 n_2 \kappa^2 L \log\frac{R}{\xi}.
\qquad
\label{eq:E:3}
\eeqn
The like-charged vortices repel each other while the vortices with opposite vorticities attract to each other.

Finishing this section, we notice that in order to get Eq.~\eq{eq:E:3} it is sufficient to take in the action~\eq{eq:S:Sigma}, instead of the single-vortex current~\eq{eq:example}, the following expression:
\beqn
\Sigma_{\alpha \beta}(x) & = & 
(\delta_{\alpha,0} \delta_{\beta,3} - \delta_{\alpha,3} \delta_{\beta,0}) \delta(x_1) 
\label{eq:example:2} \\
& & \times  [n_1 \delta(x_2 - R/2) + n_2 \delta(x_2 - R/2)]\,.
\nonumber
\eeqn

\subsection{Two-component superfluids in long-range limit}

\subsubsection{Lagrangians in the long-range limit}

The long-range limit of both Hermitian~\eq{eq:L:H} and non-Hermitian~\eq{eq:L:prime} two-condensate models can be reached by expressing the diagonal masses $m_a^2 = - 2 \lambda_a \kappa_a^2$ via the \add{condensate densities} $\kappa_a^2 > 0$ and $a = 1,2$ and then taking the limit of strong quartic interaction, $\lambda_1 = \lambda_2 \to \infty$. The parameters $\kappa_a > 0$ fix the radial amplitudes for each field, $\phi_a = \kappa_a e^{i \theta_a}$, while leaving the phases $\theta_a$ as the only dynamical variables. 
In the long-range limit, the Lagrangians for the Hermitian~\eq{eq:L:H} and non-Hermitian~\eq{eq:L:prime} models reduce, respectively, to the following expressions.
\beqn
{\cal L}_\HH & = & \phantom{+} \kappa_1^2 \partial_\nu\theta_1 \partial^\nu \theta_1 + \kappa_2^2  \partial_\nu \theta_2 \partial^\nu\theta_2 \nonumber \\
& & \hskip 15mm - 2 m_5^2 \kappa_1 \kappa_2 \cos(\theta_1 - \theta_2)\,,
\label{eq:L:H:London}
\\
{\cal L}_\NH & = & \phantom{+} \kappa_1^2 \partial_\nu\theta_1 \partial^\nu \theta_1 + \kappa_2^2  \partial_\nu \theta_2 \partial^\nu\theta_2  \nonumber \\
& & \hskip 15mm + 2 i m_5^2 \kappa_1 \kappa_2 \sin(\theta_1 - \theta_2)\,.
\label{eq:L:NH:London}
\eeqn
The only difference between the Hermitian and non-Hermitian cases appears in the interaction between the phases of different condensates: instead of the cosine function in the Hermitian model, its non-Hermitian model has a sine function preceded by a purely imaginary coupling. 

According to the third criterion of Eq.~\eq{eq:stability:NH}, the non-Hermitian theory in the long-range limit at $\kappa_1 \neq \kappa_2$ would correspond to the $\PT$ unbroken phase if the theory were non-interacting. Below, we will see that in the interacting theory this criterion, unsurprisingly, does not work. Still, Eq.~\eq{eq:L:NH:London} represents a meaningful theory even if $m_5 \neq 0$.

\subsubsection{Hermitian two-condensate model}
\label{sec:Hermitian:vortices}

\add{
The system described by the Hermitian Lagrangian~\eq{eq:L:H:London} corresponds to a long-range description of two coupled condensates within the Gross-Pitaevskii formalism~\cite{ref:Pitaevskii:book}. The condensates correspond to two different hyperfine spin states of (for example, of ${}^{87}$Rb), which are driven by the Rabi frequency $\Omega$.
}

\add{
In the static limit, the energy functional of the system~\eq{eq:L:H:London} can be written in a suggestive non-relativistic form:
\beqn
E[\theta_1,\theta_1] & = & \int d^3 x \, \biggl\{ \frac{\hbar^2}{2 m} \left[\rho_1 ({\bs \nabla} \theta_1)^2 
+ \rho_2 ({\bs \nabla} \theta_2)^2   \right] \nonumber \\
& &  - \hbar \Omega \sqrt{\rho_1 \rho_2} \cos(\theta_1 - \theta_2)  \biggr\}
\label{eq:GP}
\eeqn
where $\rho_a \equiv 2 m \kappa_a^2$ is the density of the $a=1,2$ condensate and the interaction term $\Omega \equiv - 2 m_5^2 \kappa_1 \kappa_2$ has the meaning of the Rabi frequency (here we restored for a moment the powers of $\hbar$). Below we continue to work in the notations of Eq.~\eq{eq:L:H:London}.
}

From the classical equations of motion of the Hermitian model,
\beqn
\kappa_1 \Box\, \theta_1 - m_5^2 \kappa_2 \sin(\theta_1 - \theta_2) & = & 0\,, \\
\kappa_2 \Box\, \theta_2 + m_5^2 \kappa_1\sin(\theta_1 - \theta_2) & = & 0\,,
\eeqn
one immediately determines the presence of the Goldstone massless mode 
\beqn
\chi = \theta_1 \sin^2 \beta + \theta_2 \cos^2 \beta\,,
\label{eq:chi:H}
\eeqn
and the massive excitation,
\beqn
\gamma =\theta_1 - \theta_2 \,.
\label{eq:gamma:H}
\eeqn

These degrees of freedom satisfy, respectively, the following equations:
\beqn
\Box\, \chi & = & 0\,, 
\label{eq:eos:chi}
\\
\Box\, \gamma - M^2 \sin \gamma & = & 0\,.
\label{eq:eos:gamma}
\eeqn
Here 
\beqn
M^2 = \frac{2 m_5^2}{\sin 2 \beta} \equiv \frac{\kappa_1^2 + \kappa_2^2}{\kappa_1 \kappa_2} m_5^2 \,,
\label{eq:M}
\eeqn
is the mass of the mode~\eq{eq:gamma:H}, 
\beqn
\tan \beta = \frac{\kappa_1}{\kappa_2}\,,
\eeqn
is the angle which determines the relative strength of the condensates, and 
\beqn
\kappa^2 = \kappa_1^2 + \kappa_2^2\,,
\label{eq:kappa:common}
\eeqn
\add{is the total condensate density given by the sum of the densities of the individual components.}

\add{The mode $\chi$ corresponds to the gapless sound excitation (the density wave) while mode $\gamma$ is the spin density wave which is always gapped provided the Rabi frequency is nonzero, $\Omega \neq 0$ (or, in our notations, $m_5 \neq 0$)~\cite{ref:E}.}

Since $\kappa_a > 0$, the mass~\eq{eq:M} can always be chosen as a positive quantity. The Hermitian Lagrangian~\eq{eq:L:H:London} can therefore be rewritten as a sum of independent contributions coming from massless~\eq{eq:chi:H} and massive~\eq{eq:gamma:H} fields:
\beqn
{\cal L}_\HH = \kappa^2 \partial_\nu\chi \partial^\nu \chi + {\cal L}^{(M)}_\HH(\gamma)\,.
\label{eq:L:H:London:2}
\eeqn

The Goldstone mode $\chi$ corresponds to the massless excitation of the one-component Bose gas superfluid~\eq{eq:S:XY}. The massive mode $\gamma$ is described by the sine-Gordon Lagrangian: 
\beqn
{\cal L}^{(M)}_\NH(\gamma) = {\tilde \kappa}^2 \left( \partial_\nu \gamma \partial^\nu\gamma - 2 M^2 \cos \gamma \right)\,,
\label{eq:sine:Gordon:H}
\eeqn
where the parameter
\beqn
{\tilde \kappa} = \frac{\kappa\sin 2 \beta}{2}\,,
\eeqn
plays a role of the amplitude corresponding to the massive condensate. In Eq.~\eq{eq:sine:Gordon:H}, the term $\cos\gamma$ appears naturally instead of the usual $\gamma^2$ mass term in agreement with the $\gamma \to \gamma + 2 \pi$ periodicity coming from the symmetry of the original fields $\theta_1$ and $\theta_2$. It is instructive to look at the limit of large mass, $M \xi \gg 1$, which reduces the fluctuations of the field $\gamma$ and makes the last term in the model~\eq{eq:sine:Gordon:H} quadratic. Then the integration over the field $\gamma$ can be done explicitly taking into account the vortex singularities. 

Following the analogy of the one-component model, we conclude the two-component model should contain two types of superfluid vortices associated with singularities (winding) in the phases of the fields $\theta_1$ and $\theta_2$. Using the decompositions~\eq{eq:chi:H} and~\eq{eq:gamma:H}, we identity the massless $\Sigma^{(0)}$ and massive $\Sigma^{(M)}$ combinations of the vortex worldsheets which appear as the singularities in $\chi$ and $\gamma$ fields, respectively:
\beqn
\Sigma^{(0)}_{\mu\nu} & = & \Sigma^{(1)}_{\mu\nu} \sin^2\beta + \Sigma^{(2)}_{\mu\nu} \cos^2\beta\,,
\label{eq:Sigma:0}\\
\Sigma^{(M)}_{\mu\nu} & = & \Sigma^{(1)}_{\mu\nu} - \Sigma^{(2)}_{\mu\nu}\,.
\label{eq:Sigma:M}
\eeqn
Here the individual phase windings are defined according to Eq.~\eq{eq:theta:sing}:
\beqn
\partial_{[\mu,} \partial_{\nu]} \theta^s_i (x, \tilde x) = 2 \pi \cdot \frac{1}{2} \epsilon_{\mu\nu \alpha \beta} \Sigma^{(i)}_{\alpha \beta}(x, \tilde x), \quad i = 1,2\,. \quad
\label{eq:Sigmas}
\eeqn

Therefore, the effective theory of vortices is written as follows:
\beqn
S[\Sigma] & = & \phantom{+} 4 \pi^2 \kappa^2 \int_{\Sigma^{(0)}} d^2 \sigma(\tilde x) \int_{\Sigma^{(0)}} d^2 \sigma({\tilde x}') D({\tilde x} - {\tilde x}')
\label{eq:S:Sigma:H} \\
& & + 4 \pi^2 {\tilde \kappa}^2 \int_{\Sigma^{(M)}} d^2 \sigma(\tilde x) \int_{\Sigma^{(M)}} d^2 \sigma({\tilde x}') D_M({\tilde x} - {\tilde x}')\,.
\nonumber
\eeqn
where $D_M(x)$ is the advanced Green's function corresponding to the propagator of the massive particle, $(\Box + M^2)\, D_M(x) = - \delta^{(4)}(x)$.

\add{
Despite its cumbersome appearance, the physical sense of the effective action~\eq{eq:S:Sigma:H} is relatively transparent: the vortex worldsheets interact with each other via the massless density waves and the massive spin-wave exchanges in the combinations $\Sigma^{(0)}$ and $\Sigma^{(M)}$. The interaction is given by combining the attractive long-range (Coulomb-like) potential and the short-range (Yukawa) interaction with the strength determined by the couplings $\kappa$ and $\tilde\kappa$, and the angle $\beta$ which determines the relative strength of the condensates. 
}

\add{
Let us consider what the consequence of the action~\eq{eq:S:Sigma:H} for two static parallel straight vortices of the length $L \ll \xi$ is. The Coulomb interaction of the first term in Eq.~\eq{eq:S:Sigma:H} is given by the long-range logarithmic potential~\eq{eq:E:3}. The short-range part which appears in the second term of Eq.~\eq{eq:S:Sigma:H} corresponds to the interaction (we neglect the energy of the single vortex):
\beqn
V_M(R) = - 4 \pi n_1 n_2 \kappa^2 L K_0 \left( \frac{R}{\xi} \right).
\qquad
\label{eq:Yukawa}
\eeqn
where $K_0$ is the modified Bessel function of the second kind. In the limit of zero Rabi frequency, $\Omega = 0$, the coupling between the vortices in different condensates disappears, $m_5 = 0$, the mass \eq{eq:M} vanishes, and the massive spin-wave exchange~\eq{eq:Yukawa} takes the logarithmic form~\eq{eq:E:3}. However, since the action~\eq{eq:S:Sigma:H} is derived in the limit of the massive spin-wave, $M \xi \ll 1$, the last term leads only to the short-range interactions. The logarithmic potential dominates the large-distance physics leading to the following potential between the vortices of the kinds $a,b=1,2$:
\beqn
V_{ab} (R) = - 4 \pi n_a n_b \alpha_a \alpha_b \kappa^2 L \log \left( \frac{|{\bs x}_a - {\bs x}_b|}{\xi} \right)\,,
\label{eq:Vab}
\eeqn 
where $\alpha_1 = \sin^2\beta$ and $\alpha_2 = \cos^2\beta$ gives the amplitudes of, respectively, the fields $\theta_1$ and $\theta_2$ in the massless density wave~\eq{eq:chi:H}. The expression~\eq{eq:Vab} is a trivial generalization of the vortex-vortex interaction in the case of a single condensate~\eq{eq:E:3}.
}

\add{
The two-component model possesses, however, a new object, the domain wall in the scalar condensates.~\cite{ref:E,ref:E:prime} This nontrivial topological structure has no analogue in the one-component model. Technically, its appearance can be seen from the last  (cosine) term in the two-field Lagrangian~\eq{eq:L:H:London} which leads to a soliton in the massive field $\gamma$, Eq.~\eq{eq:gamma:H}. The domain wall appears due to the coupling $m_5$, which, as we mentioned, corresponds to the Rabi frequency $\Omega$ that couples the two spin states. The field $\gamma$ changes by $2\pi$ across the vortex, which guarantees the topological stability of the wall within the model~\eq{eq:L:H:London}.
}

\add{
The boundary of a domain wall is necessarily a vortex that carries an integer winding number. Since the domain wall possesses a nonzero tension, the vortex cannot exist as an isolated object because it would otherwise have infinite energy due to the wall attached. Therefore, the vortices from different condensates are bound by a linear potential because the domain wall possesses a constant energy per unit area. Such bounded vortex pairs are sometimes called merons~\cite{ref:vortices:review}.
}

\add{
The existence of the domain wall provides us with yet another exciting link with the physics of strong interactions described by Quantum Chromodynamics (QCD). If the domain wall in the mentioned vortex pair gets broken, new vortices appear from the condensate. The created vortices produce additional vortex pairs that are attached to the broken ends of the domain wall~\cite{ref:E,ref:F}. The process is a direct analog of the confining QCD string breaking, which extends between the (anti)quarks and gets broken if a sufficiently long distance separates the quarks. Thus, the appearance of the domain wall leads to a rather nontrivial vortex-pair dynamics~\cite{ref:F,ref:G}.
}

\add{The confinement picture works only for a relatively small inter-condensate coupling given by the Rabi frequency $\Omega$ (or, for small off-diagonal mass $m_5$ in our notations). At a high inter-condensate coupling, the domain wall becomes unstable, and the vortex pairs are no more confined~\cite{ref:E}. The inter-vortex potential is then given by the long-range logarithmic potential~\eq{eq:Vab} determined by the density wave exchange.
}

\subsubsection{Non-Hermitian two-condensate model}
\label{sec:nonHermitian:vortices}

\add{
The energy of the non-Hermitian two-condensate model~\eq{eq:L:NH:London} can also be rewritten as a formal expression similar to its Hermitian counterpart~\eq{eq:GP}. In the non-Hermitian case, the meaning of the Rabi frequency $\Omega$ becomes rather transparent: the Rabi coupling $\Omega$ should couple to the different hyperfine spin components with a different sign. Leaving aside the physical realization of such scenario, we notice that the non-Hermitian case can be analyzed closely paralleling the Hermitian case considered above. In particular,
} 
exactly the same field combinations~\eq{eq:chi:H} and \eq{eq:gamma:H} can be used to rewrite the non-Hermitian theory~\eq{eq:L:NH:London} in terms of the massless (density wave) and massive (spin wave) phase combinations:
\beqn
{\cal L}_\NH = \kappa^2 \partial_\nu\chi \partial^\nu \chi + {\cal L}^{(M)}_\NH(\gamma)\,.
\label{eq:L:NH:London:2}
\eeqn
This non-Hermitian model possesses the usual Hermitian Goldstone mode $\chi$ which leads to the long-range interactions between the combinations of the worldsheets~\eq{eq:Sigma:0}. However, the would-be massive excitation $\gamma$ exhibits a non-Hermitian behavior described by the Lagrangian:
\beqn
{\cal L}^{(M)}_\NH(\gamma) = {\tilde \kappa}^2 \left( \partial_\nu \gamma \partial^\nu\gamma + 2 i M^2 \sin \gamma \right)\,,
\label{eq:sine:Gordon:NH}
\eeqn
and therefore the interaction between the would-be massive components of the vortex sheets~\eq{eq:Sigma:M} is not evident.

The model~\eq{eq:sine:Gordon:NH} is nothing but a non-Hermitian version of the Sine-Gordon model 
in (3+1) dimensions. As a side remark, we notice that the Lagrangian~\eq{eq:sine:Gordon:NH} appears as a bosonic dual of the non-Hermitian massive Thirring Model in (1+1) dimensions~\cite{Bender:2005hf}. According to Ref.~\cite{Bender:2005hf}, this model with the purely imaginary coupling in front of the sine-term resides in the $\PT$-broken domain and, therefore, should be characterized by complex energy dispersions which correspond to dissipation or instability, or the both. 

The second term in the Lagrangian~\eq{eq:sine:Gordon:NH} implies that the $\gamma=0$ point is not a local extremum of the corresponding action. The stable minima could appear around the values $\gamma_\pm = \pm \pi/2$. Defining $\gamma = \pm \pi/2 + \delta \gamma$, we get the following equations of motion:
\beqn
\Box\, \delta \gamma \pm i M^2 \sin \delta\gamma = 0\,,
\label{eq:eos:gamma:NH}
\eeqn
which differ from the classical equations of motion of the Hermitian model~\eq{eq:eos:gamma} by the purely complex coefficient in front of the sine term. For small fluctuations around the minimum, $\delta\gamma = 0$, the solutions of Eq.~\eq{eq:eos:gamma:NH} give us the dispersions for the energy: $\omega_{\bs k} = \sqrt{{\bs k}^2 \pm i M^2}$. We find that the particle-like, positive-energy solutions with $\mathrm{Re}\,\omega>0$ lead to an explosive behavior near the $\gamma = - \pi/2$ minimum which appears to be unstable. For example, the amplitude of any zero-momentum solution (${\bs k} = 0$) diverges with time $t$ as $\delta \gamma \sim \exp(\Gamma t)$ where 
\beqn
\Gamma = \frac{M}{\sqrt{2}} \equiv \frac{\kappa_1^2 + \kappa_2^2}{\kappa_1 \kappa_2} \frac{m_5^2}{\sqrt{2}} > 0\,.
\label{eq:Gamma}
\eeqn
In the language of a Hermitian theory, the point $\gamma = - \pi/2$ would correspond to an extremum.

However, the minimum $\gamma = + \pi/2$ is stable so that all particle excitations around it behave as dissipative solutions $\delta \gamma \sim \exp(- \Gamma t)$ that approach the minimum point $\gamma = + \pi/2$ should the field $\gamma$ deviate from it. Since the angle $\gamma$ takes a constant value in the stable minimum, it evidently means that the field $\gamma$ contains no vortex singularities in this ground state. According to Eq.~\eq{eq:Sigma:M}, the vortex singularities in the both phases $\theta_1$ and $\theta_2$ should coincide with each other: $\Sigma^{(M)}_{\mu\nu} = \Sigma^{(1)}_{\mu\nu} - \Sigma^{(2)}_{\mu\nu} = 0$. 

Thus, the vortices in the $\phi_1$ and $\phi_2$ condensate can only exist provided they coincide with each other, $\Sigma^{(0)}_{\mu\nu} = \Sigma^{(1)}_{\mu\nu} = \Sigma^{(2)}_{\mu\nu}$ thus forming a single double-vortex sheet $\Sigma^{(0)}_{\mu\nu}$ according to Eq.~\eq{eq:Sigma:0}. Any fluctuation that separates the vortices leads to the energy dissipation with the dissipation rate~\eq{eq:Gamma} which returns the vortices back to their common stable non-dissipative minimum. 

The common vortex line is described by action of the vortex in the one-component condensate~\eq{eq:S:Sigma}, where the coupling $\kappa$ is given in Eq.~\eq{eq:kappa:common}. The energy per unit length of the joint vortex is thus given by Eq.~\eq{eq:E:1}. In this state, the common vortex segments interact with other segments via a long-range interaction mediated by massless particles. For the straight static vortices separated by the distance $R$, the interaction is given by Eq.~\eq{eq:E:3}.

\add{
As we mentioned earlier, the Hermitian counterpart~\eq{eq:L:H} of the non-Hermitian system~\eq{eq:L:prime} appears in various physical contexts, including color superconductors in high-density quark matter described by QCD, relevant for the description of, for example, the interior of the neutron stars~\cite{ref:A,ref:B}, and the axion cosmic strings~\cite{ref:C}. In the field-theoretical context, both original U(1) degrees of freedom, representing the phases of the scalar condensates, mix up to form the U(1)${}_B$ (``baryon'') global symmetry where both condensates transform with the same phase and U(1)${}_A$ ("axial") global symmetry which rotates the phases of the fields oppositely.}\footnote{
\add{The terms are taken from chiral physics of (massless) Dirac fermions where the vector and chiral rotations in internal space transform upper (right-handed) and lower (left-handed) Weyl components with the same (opposite) phases which corresponds to the vector (baryon) and axial rotations, respectively~\cite{peskin95}.}
} 
\add{In our models, these global symmetries correspond to the Goldstone massless mode~\eq{eq:chi:H} and the massive excitation~\eq{eq:gamma:H}, respectively.
}

\add{In particle physics, the axial symmetry is usually broken either by the axial quantum anomaly~\cite{Shifman:1988zk} or explicitly, by the (strange) quark mass, depending on the physical system described by the model. This effect is important in the cosmological context as it is responsible for the rapid decay of axion cosmic strings~\cite{ref:C}. These topological defects, which are suggested to be formed in early moments of our Universe, gets bounded to the QCD domain walls at the QCD finite-temperature phase transition. The walls decay rapidly after the QCD phase transition thus leaving no trace of the cosmic strings in the present-day Universe~\cite{ref:D}. We will not pursue this topic further concentrating, instead, on the properties of vortices themselves in the contexts of the models. We notice that in model~\eq{eq:L:H}, the ``axial'' invariance is explicitly violated by the nonzero coupling $m_5 \neq 0$ between the two fields in the scalar doublet which makes the overall mechanism similar to the explicit breaking of the axial symmetry in QCD.}

\subsubsection{Vortices in Hermitian and non-Hermitian models in long-range limit: a brief comparison}

\add{
It is worth comparing the properties of the vortices in Hermitian and non-Hermitian versions of the two-field model in the long-range limit. We have already seen that outside the long-range limit, the phase diagram of the Hermitian model (Section ~\ref{sec:Hermitian:ground:state}) has a much simpler structure as compared to the rich and complicated diagram which characterizes the phases of the non-Hermitian model (Section~\ref{sec:nonHermitian:ground:state}). The properties of the corresponding vortices, discussed in Sections~\ref{sec:Hermitian:vortices} and \ref{sec:nonHermitian:vortices}, respectively, differ dramatically as well. 
}

\add{
The Hermitian version of the model possesses two types of stable vortices associated with the phases of both condensates~\eq{eq:Sigmas}. The vortices interact with each other via nonlocal action~\eq{eq:S:Sigma:H} which describes the massive and massless interaction of the vortex worldsheets via exchanges of the massless (density wave) and massive (spin wave) excitations. The model also possesses the domain wall, a coherent structure in both condensates. However, in the limit of the sizeable inter-condensate coupling (given by the Rabi coupling $\Omega$ or the off-diagonal mass $m_5$, in our notations), the domain wall becomes unstable~\cite{ref:E} and the long-range vortex dynamics is determined by the repulsive logarithmic potential.
}

\add{
The long-range limit of the non-Hermitian model resides in the $\PT$-broken phase, which indicates, according to the general arguments discussed earlier, instability. The instability indeed appears as a dissipation provided the vortex cores of different types are not overlapping exactly. Therefore, the separated vortices from different condensates tend to attract each other until they form a tightly bound pair that is perfectly stable. 
}

\add{
Therefore, the behavior of the vortices in the Hermitian and non-Hermitian cases are somewhat similar: both of them tend to form pairs, albeit for different reasons. However, one should notice that the similarity is not exact: the binding of vortices in the Hermitian model appears due to the domain wall, which exists at low inter-condensate coupling, while the dissipative vortex binding in the non-Hermitian case is realized at the large values of this coupling. 
}

\section{Vortices in general case}
\label{sec:vortices:finite}

\add{The non-Hermitian two-field model possesses vortex solutions not only in the London (long-range) limit, where the amplitudes are fixed by the large quartic couplings $\lambda_1$ and $\lambda_2$ but also at finite values of the quartic couplings. The general is particularly interesting for the properties of the system at the scales of the order or smaller than the healing length.}

In this section, we consider examples of the static straight vortex solutions of the classical equations of motion~\eq{eq:eqns:motion} assuming the standard axial ansatz for the scalar fields 
\beqn
\phi_a(r,\theta) = v_a f_a(r) e^{i n \theta}\,, \qquad a = 1,2, \quad
\label{eq:phi:anzats}
\eeqn
where $r$ and $\theta$ are the radial coordinates in the $(x_1,x_2)$ plane and $n \in \Z$ is the vorticity of the solution. These static and straight field configurations do not depend on time and $x_3$ coordinates. The vacuum values of the condensates, $v_1$ and $v_2$, are the solutions of Eqs.~\eq{eq:class:2}.

The consistency of the coupled solutions at $m_5 \neq 0$ imply that the vortices in $\phi_1$ and $\phi_2$ condensates should be superimposed on each other and they should have the same winding numbers $n_1 = n_2 = n$ in Eq.~\eq{eq:phi:anzats}. Below we concentrate on the non-Hermitian model which is the subject of our paper (the analysis of the Hermitian counterpart can also be performed in the same way). 

The radial profiles of the vortices are described by the functions $f_a$ with the following asymptotics:
\beqn
\lim_{r \to \infty} f_a(r) = 1, 
\qquad  
\lim_{r \to 0} f_a(r) = 0, \qquad a = 1,2, \qquad
\label{eq:asymptotics}
\eeqn
which guarantee that the total energy of the vortex solution is converging both at the spatial infinity and at the origin, respectively. 

The classical equations of motion~\eq{eq:eqns:motion} lead to the following system of equations for the profile functions:
\beqs
\beqn
f_1''(r) + \frac{f_1'(r)}{r} - \frac{n^2}{r^2}f_1(r) - m_1^2 f_1(r) - m_5^2 \frac{v_2}{v_1} f_2(r) & &  \\
- 2\lambda_1 v_1^2 f_1^3 & = & 0\,,
\nonumber \\
f_2''(r) + \frac{f_2'(r)}{r} - \frac{n^2}{r^2}f_2(r) - m_2^2 f_2(r) + m_5^2 \frac{v_1}{v_2} f_1(r) & & \\
- 2\lambda_2 v_2^2 f_2^3 & = & 0\,,
\nonumber 
\eeqn
\label{eq:eqs:f:s}
\eeqs
which, evidently, do not possess straightforward analytical solutions. 

Close to the origin, the last non-linear terms in both equations~\eq{eq:eqs:f:s} can be neglected and the differential equations can be linearized. The solutions can be represented in the form of the polynomials,
\beqn
f_a(r) = \sum_{k=0}^\infty A_a^{(2 k)} r^{n + 2 k}\,,
\label{eq:f:a}
\eeqn
which involve only the positive even powers of the radius starting from the power $r^n$ determined by the vorticity number $n = 1,2,\dots$. The ansatz~\eq{eq:f:a} is thus consistent with the asymptotics~\eq{eq:asymptotics}. One gets for the first three coefficients:
\beqs
\beqn 
A_1^{(0)} & = & A_1\,, \qquad A_1^{(2)} = \frac{A_1 v_1 m_1^2 + A_2 v_2 m_5^2}{4(n+1) v_1}\,,
\nonumber \\
A_1^{(4)} & = & \frac{A_1 v_1 (m_1^4 - m_5^4) + A_2 v_2 m_5^2 (m_1^2+m_2^2)}{32(n+1)(n+2) v_1}\,, \qquad 
\label{eq:An}\\
A_2^{(0)} & = & A_2\,, \qquad A_2^{(2)} = \frac{A_2 v_2 m_1^2 - A_1 v_1 m_5^2}{4(n+1) v_2}\,,
\nonumber \\
A_2^{(4)} & = & \frac{A_2 v_2 (m_1^4 - m_5^4) - A_1 v_1 m_5^2 (m_1^2+m_2^2)}{32(n+1)(n+2) v_2}\,, \qquad
\nonumber
\eeqn
\label{eq:An:s}
\eeqs
where $A_1$ and $A_2$ are free parameters of the solution which cannot be fixed at this stage. The series~\eq{eq:An} of the $f_1$ and $f_2$ profile functions are related to each other by the flip of the sign in front of the off-diagonal mass term $m_5^2$.

In the large-distance region, $r \to \infty$, the asymptotics~\eq{eq:asymptotics} imply $f_a = 1 - h_a$ where $|h_a| \ll 1$ at sufficiently large distances. The linearized equations of motion,
\beqn
h_1''(r) + \frac{h_1'(r)}{r} - \frac{n^2}{r^2} h_1(r)- m_1^2 h_1(r) - m_5^2\frac{v_2}{v_1}h_2(r) & &  \\
- 6\lambda_1 v_1^2 h_1(r) & = & 0\,, \nonumber \\
h_2''(r) + \frac{h_2'(r)}{r} - \frac{n^2}{r^2} h_2(r)- m_2^2 h_2(r) + m_5^2\frac{v_1}{v_2}h_1(r) & & \\
- 6\lambda_2 v_2^2 h_2(r) & = & 0\,, \nonumber 
\eeqn
suggest that their solutions can be represented in the following form:
\begin{equation}
h_a(r) = B_a^{(0)} r^s e^{- \mu r}\,.
\label{eq:h:a}
\end{equation}
The self-consistency of the solutions provides us with the power $s = - 1/2$ of the algebraic prefactor $r^s$ and also imposes two simultaneous constraints:
\beqs
\beqn
\mu^2 - m_1^2 - m_5^2 \frac{B_2^{(0)} v_2}{B_1^{(0)} v_1} - 6\lambda_1 v_1^2 & = & 0\,,\\
\mu^2 - m_2^2 + m_5^2 \frac{B_1^{(0)} v_1}{B_2^{(0)} v_2} - 6\lambda_2 v_2^2 & = & 0\,.
\eeqn
\label{eq:eqs:mu}
\eeqs
These equations give us two possible solutions for the ratio of the coefficients $B_a^{(0)}$ from Eq.~\eq{eq:h:a}:
\beqn
\frac{B_1^{(0)}}{B_2^{(0)}} & = & \frac{v_2}{v_1} \left( \alpha \pm \sqrt{\alpha^2 - 1}  \right)\,,
\eeqn
and also determines the common exponent:
\beqn
\mu = \sqrt{ 6 \lambda_1 v_1^2 + m_1^2 + m_5^2  \left( \alpha \pm \sqrt{\alpha^2 - 1}  \right)}\,.
\eeqn
We denoted for brevity:
\beqn
\alpha = \frac{6 (\lambda_2 v^2_2 - \lambda_1 v^2_1) + m_2^2 - m_1^2}{2 m_5^2}\,.
\eeqn

The next-order correction to Eq.~\eq{eq:h:a} can also be easily obtained,
\begin{equation}
    h_a(r)=\left(\frac{B_a^{(0)}}{r^{\frac{1}{2}}} + \frac{B^{(1)}_a}{r^{\frac{3}{2}}}+ O\left( r^{-\frac{5}{2}} \right) \right)e^{-\mu r}\,,
\label{eq:h:a:r}
\end{equation}
where
\begin{equation}
    B_a^{(1)} = \frac{1}{2\mu}\left(n^2-\frac{1}{4}\right) B_a^{(0)}\,.
\label{eq:B:a:r}
\end{equation}

\begin{figure}[!htb]
\center{\includegraphics[scale=0.375]{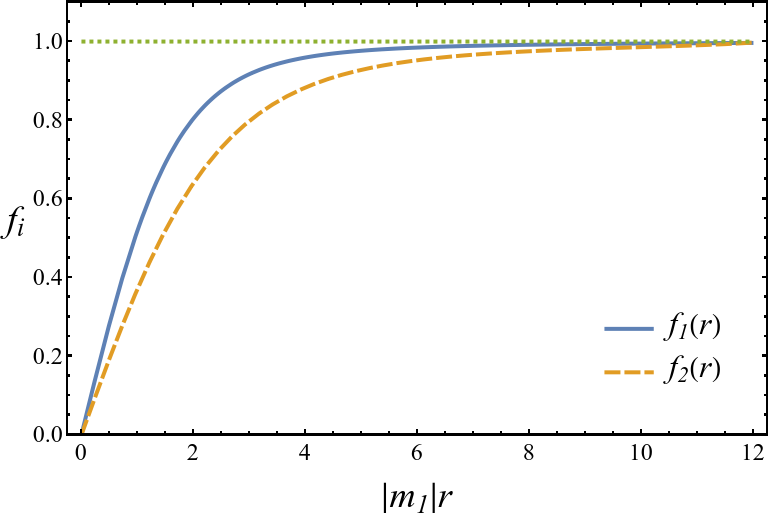}}
\caption{The profile functions of the elementary $n=1$ vortex solution at the mass parameters $m_2^2= |m_1^2|$ and $m_5^2= 0.1 |m_1^2|$ with $m_1^2 < 0$ and the equal quartic couplings $\lambda_1 = \lambda_2 = 1$.}
\label{fig:profile:functions}
\end{figure}

If the off-diagonal mass vanishes, $m_5^2 = 0$, the non-Hermitian two-scalar model reduces to two non-interacting scalar models 
${\cal L}(\phi_1,\phi_2) = {\cal L}_1(\phi_1) + {\cal L}_2(\phi_2)$ with
\begin{equation}
{\cal L}_a(\phi_a) = \partial_\nu\phi^*_a \partial^\nu\phi_a - m^2_a|\phi_a|^2 - \lambda_a\phi^4_a.
\end{equation}
This single-field model possesses the asymptotic solutions of the form (we omit the index $a$ for simplicity):
\beqn
f(r) & = & A \left[ r^n+\frac{m^2}{4(n+1)} r^{n+2} \right. \nonumber \\
& & \left. + \frac{m^4}{32(n+1)(n+2)} r^{n+4} + O\left(r^{n+6}\right) \right] \,,
\label{eq:asympt:one:field}\\
h(r) & = & B \left[ \frac{1}{\sqrt{r}} + \frac{1}{2\mu r^{3/2}} \left(n^2-\frac{1}{4}\right) + O\left( r^{-5/2} \right) \right] \, e^{-\mu r} \,,
\nonumber
\eeqn
where the asymptotic behavior is controlled by the mass of the single scalar field:
\begin{equation}
    \mu = \sqrt{m^2+6\lambda v^2}=2\sqrt{\lambda}v^2\,.
\label{eq:mu:sol}
\end{equation}
In this limit, the two equations~\eq{eq:eqs:mu} decouple, and the mass parameters reduce to Eq.~\eq{eq:mu:sol} for each field. The asymptotics of the non-Hermitian solution~\eq{eq:f:a}, \eq{eq:An:s}, \eq{eq:h:a:r}, and \eq{eq:B:a:r} are consistent with the single-field solution~\eq{eq:asympt:one:field} as well. 

Our numerical analysis confirms the existence of the stable vortex solutions in the regions of the phase diagram with non-zero condensates. An example of the profile functions for a set of coupling constants is shown in Fig.~\ref{fig:profile:functions}. All the radial $n=1$ profiles of the vortices exhibit the same qualitative features, the linear rise close to the origin and the exponentially slow approach of the corresponding vacuum expectation values at large distances. These properties reveal the generic behavior of all solutions that we have analyzed. 

The energy density of the non-Hermitian vortex (calculated per unit vortex length),
\beqn
E_{\NH} & = & 2\pi \int_0^\infty r d r \bigl(|\nabla\phi_1|^2+|\nabla\phi_2|^2  \\
& & \hskip 20mm + m_1^2\phi_1^2+m_2^2\phi_2^2 + \lambda_1\phi_1^4+\lambda_2\phi_2^4\bigr), \qquad
\nonumber 
\eeqn
can be simplified with the use of the corresponding equations of motion~\eq{eq:eqns:motion}. It can be expressed via the profile functions $f_a$ as follows:
\begin{equation}
E = 2 \pi \int_0^\infty r d r \left\{ \lambda_1 v_1^4 [ 1 - f_1^4(r)] + \lambda_2 v_2^4 [1 - f_2^4(r)] \right\}\,,
\label{eq:E:vort}
\end{equation}
where the energy is normalized in such a way that $E = 0$ in the absence of the vortex. The very same expression~\eq{eq:E:vort} also gives us the energy of the vortex in the counterpart Hermitian theory~\eq{eq:L:H},
\beqn 
E_\HH  = 2\pi\int_0^\infty r d r \bigl( & & |\nabla\phi_1|^2 + |\nabla\phi_2|^2 + m_1^2\phi_1^2 + m_2^2\phi_2^2 \nonumber \\
& & + 2m_5^2\phi_1\phi_2 + \lambda_1\phi_1^4 + \lambda_2\phi_2^4\bigr)\,,
\eeqn 
after the use of the corresponding classical equations of motion~\eq{eq:eqns:H}.

\begin{figure}[!htb]
\includegraphics[scale=0.375]{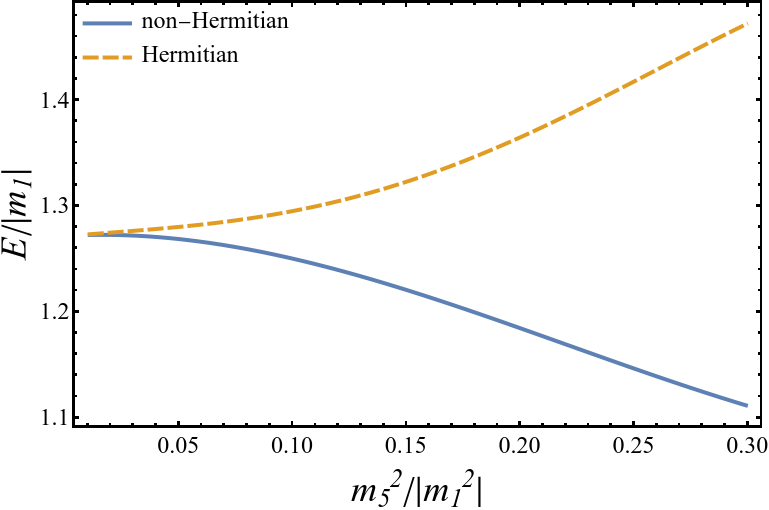}\\
(a) \\[5mm]
\includegraphics[scale=0.375]{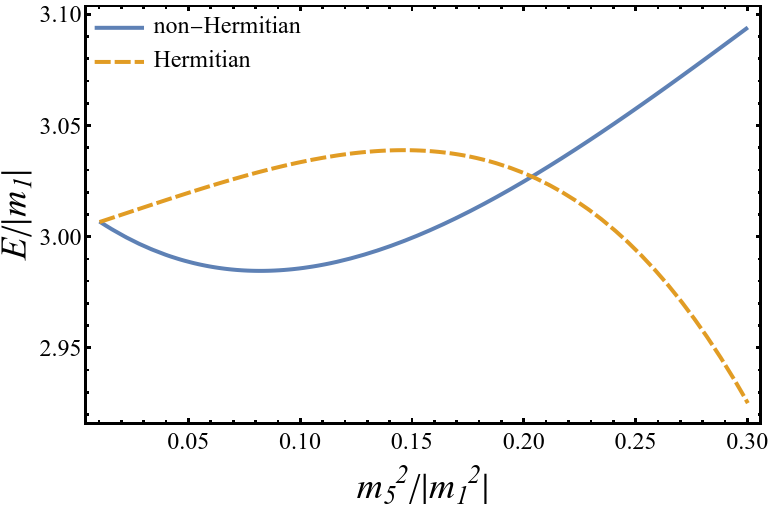} \\
(b)
\caption{Hermitian and non-Hermitian vortex energies vs. the off-diagonal mass squared $m_5^2$ in different stability areas
(a) $m_2^2 = - m_1^2 >0$ corresponding to the border of the stable and unstable regions and 
(b) $m_2^2 = 2.5 m_1^2 < 0$ residing within the stable region, at $\lambda_1 = \lambda_2=1$, Fig.~\ref{fig:stability:areas:NH}(g).}
\label{fig:vortex:energy}
\end{figure}

Finally, despite of the mundane similarity of the numerically obtained vortex configurations in various regions, one notices rather unusual difference of the evolution of the vortex energy as the function of the off-diagonal mass parameter $m_5$ in Hermitian and non-Hermitian regions. We show the examples of vortex energies in Hermitian and non-Hermitian theories in different stability areas, with $m_1^2 <0$ and $m_2^2 > 0$ in Fig.~\ref{fig:vortex:energy}(a) and with $m_1^2, m_2^2 < 0$ in Fig.~\ref{fig:vortex:energy}(b) at the same values of quartic couplings. The vortex energies in the Hermitian and non-Hermitian versions trivially coincide at $m_5 = 0$ and then they tend to separate as the off-diagonal mass $m_5$ increases.  One notices non-monotonic behavior of energies in the completely broken ($m_1^2, m_2^2 < 0$) part of the phase diagram.

\section{Discussion and Conclusion}

\add{
Our paper discusses the phase diagram and reveals the properties of line-like topological excitations and vortices in the non-Hermitian relativistic model of two interacting scalar complex fields. We concentrate on non-Hermitian parity-time ($\PT$) symmetric realization of the model, which describes an open system that communicates with an external environment with precisely balanced gains and losses. The exact balance is encoded in the $\PT$ symmetry of the model Lagrangian. This symmetry ensures the stability of its steady ground state provided that the $\PT$-invariance of the model is not broken spontaneously. 
}

\add{
The model describes the properties of two scalar condensates that can exhibit the spontaneous breaking of the discrete $\PT$ symmetry in addition to the spontaneous breaking of the continuous global U(1) symmetry. Given the presence of the two symmetries that can be either broken or unbroken, the model thus contains four different phases, which makes the phase structure of the model very nontrivial.
}

\add{
The phase diagram of the non-Hermitian two-condensate model is shown in Figs.~\ref{fig:energy:condensates} and \ref{fig:stability:areas:NH}. We compared the non-Hermitian model with its Hermitian counterpart with identical couplings (Fig.~\ref{fig:energy:condensates}), and we have observed the striking difference between the two cases. While the Hermitian model hosts the stable U(1) broken phase almost at every point of the phase diagram (except for a thin line, where the condensates decouple), the non-Hermitian diagram shows a wide variety of intertwined stable and unstable phases with both broken and respected U(1) symmetry. The richness of the phase diagram is guaranteed by the nontrivial pattern of the spontaneous $\PT$--symmetry breaking, which is generalized, in our paper, to the interacting theory.
}

\add{On the practical level, the stability of the ground state} can be identified via the absence of the negative modes in fluctuation matrix ${\cal M}^2_{\NH}$ that describes the quadratic fluctuations of the fields~\eq{eq:F:M} over the ground-state condensates. On the contrary, in the spontaneously $\PT$-broken regions, the quadratic fluctuation matrix contains at least a single negative eigenvalue. 

In addition to the phase diagram, we studied the basic properties of global vortices in the non-Hermitian model in various regimes.

Firstly, we considered the model in the limit where the lengths of the condensates are frozen and the analytical analysis is feasible. To reveal the vortex properties, we used a set of exact transformations of the field variables that did not involve the explicit solution of the equations of motion. Noticing that the model resides in the $\PT$-broken phase \add{where the vortices are expected to be unstable}, we show that the superfluid vortices can propagate non-dissipatively if and only if the vortex singularities in different condensates have the same vorticity (winding number) and, in addition, they overlap. \add{In the case of the strong inter-condensate coupling determined by the non-Hermitian mass $m_5$,} the joint vortex segments interact via a long-ranged exchange of a massless excitation, similarly to the vortices in a Hermitian one-condensate model. The dissipation rate of the individual (separated) vortex segments is controlled by the off-diagonal mass $m_5$, which, in turn, determines the interaction between the condensates.

\add{
The behavior of the vortices in the Hermitian and non-Hermitian cases is qualitatively similar: both of them are forming pairs. However, the similarity is not exact for two reasons: In the Hermitian model, the binding of vortices proceeds via a formation of the domain wall that emerges between the vortices. In the non-Hermitian case, the binding appears due to the dissipative dynamics of the vortices until they overlap. Moreover, the confining domain walls exist at low inter-condensate coupling values contrary to the non-Hermitian case in which the dissipative vortex binding is realized at the large inter-condensate coupling.
}

Secondly, we also studied the classical vortex configurations at finite quartic couplings. In order to identify the classical configurations, we used a single set of classical equations of motion, which is obtained by the variation of the action with respect to the original fields. We omitted the equivalent but the incompatible, complimentary set of equations that correspond to the action variation with respect to the conjugated fields. This procedure, which follows Refs.~\cite{Alexandre:2017foi,Alexandre:2019jdb}, seems appropriate for the open systems residing in a steady-state regime, which does not necessarily coincide with the (thermal) equilibrium. Moreover, in this approach, the classical solutions possess a real-valued energy spectrum bounded below. The latter property is essential on the practical level as we search for the classical states using a (numerical) procedure based on the energy minimization as a criterion for the true (ground) state.

An alternative approach, based on the similarity transformation, does not possess the incompatibility of the two sets of equations of motion. This property makes the analytical procedure of finding the classical solutions more elegant~\cite{Mannheim:2018dur,Fring:2019hue}. However, the same approach makes the kinetic term (of, at least, one of the fields) negatively valued, leading to the emergence of a negative quadratic mode for the classical the solutions. Therefore, the mentioned class alternative solutions correspond to different, sphaleron-type saddle-point configurations, which can be significant for the thermal properties of the system. 

We found that the $\PT$-symmetric two-component model admits the vortex solutions inside and at the border of the $\PT$-broken regions. These two-condensate vortex solutions share similar behavior with the vortices in the one-component relativistic superfluids. For consistency of the classical solution, the vortices of different condensates should have the same position in space-time and possess the same vorticity (winding number). 

\add{Finally, we proposed that a non-Hermitian two-condensate system can be realized in the Bose-Einstein condensates of atoms in two hyperfine spin states with the asymmetric Rabi coupling $\Omega$, which couples to the different spin components with a different sign.}

\acknowledgments

The authors are grateful to Andreas Fring and Peter Millington for useful discussions and valuable comments. 
The work is supported by Grant No. 0657-2020-0015 of the Ministry of Science and Higher Education of Russia.

\end{document}